\documentclass[acmtog,author,nonacm]{acmart}
\usepackage{wrapfig,graphicx}
\usepackage{xcolor}

\usepackage{amssymb}
\usepackage{amsmath}
\usepackage{amsfonts}
\usepackage{array}
\usepackage{subcaption}
\usepackage{xcolor}
\usepackage{tabularray}
\usepackage{siunitx}
\usepackage{algorithm}
\usepackage{comment}
\usepackage{svg}
\usepackage{booktabs}
\usepackage{multirow}
\usepackage{float}
\usepackage{amsmath,enumitem}

\usepackage{amsthm}
\newtheorem{theorem}{Theorem} \newtheorem{lemma}{Lemma} \newtheorem{corollary}{Corollary}

\usepackage{adjustbox}

\usepackage{algorithm,multirow}
\usepackage[noend]{algpseudocode}

\usepackage[table]{xcolor}

\definecolor{lightgray}{gray}{0.9}

\definecolor{headbg}{gray}{0.88}
\definecolor{coltint}{gray}{0.96}
\AtBeginDocument{%
  }

\copyrightyear{2026}
\acmYear{2026}

\newcommand{\authorcomment}[3]{{\color{#2} \textbf{#1} {#3}}}
\newcommand{\amal}[1]{\authorcomment{}{blue}{#1}}
\newcommand{\stef}[1]{\authorcomment{}{red}{#1}}

\begin{document}

\title{FOMOS: Feature-Oriented Meshing of SDFs}


\author{Stefan Ohrhallinger}
\email{ohrhallinger@cg.tuwien.ac.at}
\orcid{0000-0002-2526-7700}
\affiliation{%
  \institution{TU Wien}
  \city{Wien}
  \country{Austria}
}

\author{Amal Dev Parakkat}
\email{amal.parakkat@ip.paris.fr}
\orcid{0000-0002-7554-3291}
\affiliation{%
  \institution{LTCI-Telecom Paris, Institut Polytechnique de Paris}
  \city{Paris}
  \country{France}
}

\renewcommand{\shortauthors}{Ohrhallinger and Parakkat}

\begin{abstract}
We introduce FOMOS, a purely geometric method for feature-oriented meshing of signed distance functions (SDFs) that addresses various challenges of the existing state-of-the-art methods. Our algorithm improves mesh quality and runtime, enabling the SDF meshing in a matter of seconds. While existing techniques such as Marching Cubes and Neural Dual Contouring either produce dense, low-quality triangulations or sacrifice details for sparsity, often requiring carefully chosen user parameters, our single-parameter FOMOS leverages feature sized sampling to dynamically adjust mesh resolution based on local feature size (LFS), which we compute efficiently by sampling a medial scaffold. The LFS considers not only curvature but also shape thickness and captures large features through non-local structure, omitting many redundant small triangles from existing methods. Through extensive comparisons, we demonstrate the distinct strengths of FOMOS in execution time, reconstructed result accuracy, and triangle quality in the output mesh.
We also introduce a lightweight variant of FOMOS targeting huge ($10k^3$) sparse inputs, which, to the best of our knowledge, no existing method can handle except Marching Cubes. To demonstrate the usefulness of our method, we show that FOMOS enables instant previews using a sub-sampled medial scaffold and can accelerate volumetric simulations by orders of magnitude.
\textit{Code will be released upon acceptance.}
\end{abstract}


\begin{teaserfigure}
\includegraphics[width=\linewidth]{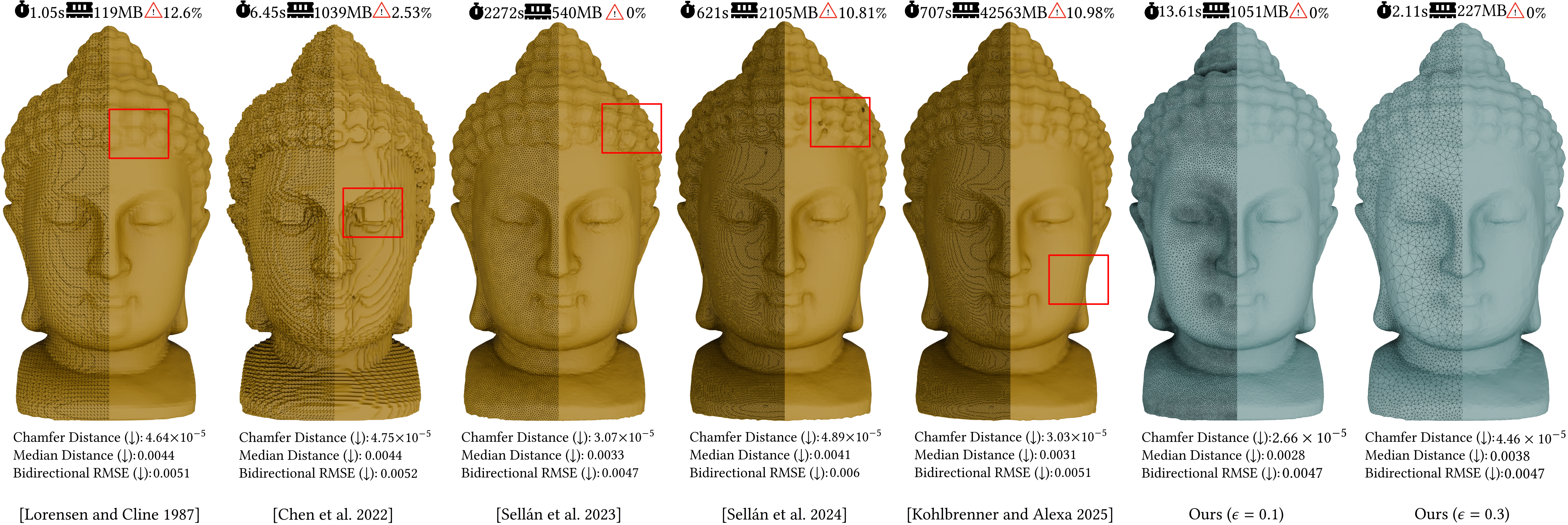}
  \caption{Comparative evaluation of SDF meshing algorithms on a $128^3$ grid showing tradeoffs between speed (1.05s-2272s), memory (119MB-42563MB), mesh quality (0.0-12.6\% defective sliver triangles, marked as triangle with exclamation), and geometric accuracy ($\num{2.81E-5}$ - $\num{4.89E-5}$ CD). From left: Marching Cubes (MC, fastest, high errors, and bad triangle quality), Neural Dual Contouring (NDC, learned priors, non-manifold and unclean result on out-of-distribution data), Reach for the Spheres (RFTS, missing details and large execution time - ran until convergence after parameter tuning), Reach for the Arcs (RFTA, poorly reconstructed regions, dense meshing, bad triangles), Maximal Empty Sphere Contouring (MESC, slow, very high memory, bad triangles), and Our FOMOS for two sampling densities (fast, less memory, no errors and nicely shaped triangles).  Red rectangles highlight reconstruction artifacts. 
  }
  \label{fig:teaser}
\end{teaserfigure}

\received{XXXX}
\received[revised]{XXXX}
\received[accepted]{XXXX}

\maketitle










\section{Introduction}



A Signed Distance Field (SDF) is a scalar function that, for any given point in space, returns the distance to the nearest point on a given boundary, with the sign indicating whether the point lies inside or outside the boundary. As a classical representation, SDFs have served as a fundamental building block for a wide range of applications in shape modeling and geometry processing. In addition to these traditional applications, recent advances in machine/deep learning have increasingly relied on SDFs for geometric deep learning, thanks to their advantages in continuous shape encoding and differentiable geometric operations. Despite the advantages of a continuous representation, which ensures resolution independence and facilitates procedural modeling and shape reconstruction (e.g., DeepSDF \cite{park2019deepsdf}), it is often discretized into triangular meshes to support various downstream applications. This discretization is typically preferred as it can be easily parallelized, seamlessly integrated into standard rendering pipelines, and efficiently stored. 

Mesh generation from SDFs is often performed by sampling the function at discrete points, typically using grids, octrees, or point clouds. Although these methods are simple and easy to implement and use, their reliance on uniform sampling prioritizes volumetric coverage over geometric fidelity. This often leads to the omission of important edges, corners, or surface curvature variations, as it does not explicitly preserve these features. Consequently, it introduces approximation errors in critical applications such as computational fluid dynamics (CFD) and finite element methods (FEM), which reduce simulation accuracy. Although it is possible to implement workarounds using a very dense grid sampling followed by feature-aware remeshing, such solutions are both computationally expensive and require significant peak memory as volumetric data grows with cubic complexity. Furthermore, such workarounds are suboptimal as they do not account for the global shape structure. 

While traditional approaches, such as Marching Cubes and Dual Contouring (along with their numerous variants), rely on grids, recent techniques have investigated grid-free meshing using learned representations and optimization techniques \cite{sellan2023reach,sellan2024reach}. However, all of these methods, as shown in Figure \ref{fig:teaser}, are prone to over-smoothing, and some impose restrictions on the topology. The focus of this work is, therefore, to automatically and precisely mesh SDFs without making any topological assumptions.
The core challenge is to develop efficient techniques for estimating features that are defined in a global sense. Our approach directly addresses the fundamental relationship between SDFs and features that must be explicitly preserved to achieve a high-fidelity reconstruction.


\textcolor{black}{Our core idea is to adopt an $\varepsilon$-sampling framework that determines the sampling density according to the local feature size (LFS), defined as the distance from a surface point to the medial axis \cite{amenta1998crust} (please see Section \ref{sec:definitions} for a formal definition). This choice addresses key limitations of conventional sampling strategies: uniform-length sampling uses a fixed spatial scale regardless of local geometric complexity, whereas curvature-aware sampling relies solely on surface curvature alone, which may not fully capture the scale of thin structures or closely spaced surface regions (particularly useful for volumetric meshing). In contrast, $\varepsilon$-sampling based on LFS accounts for both local curvature and the spatial proximity of surface regions, providing a principled basis to preserve fine geometric details while avoiding unnecessary sampling in less complex regions~\cite{amenta1998surface,boissonnat2005provably}.} To achieve this, we introduce a new technique for computing the local feature size (LFS), and subsequently incorporate an iterative refinement process to add points that meet that sampling condition, thereby preserving and capturing features of arbitrary size. We demonstrate that our method outperforms the state of the art in both quality and runtime, further enabling instant in-memory meshing of SDFs with huge grid sizes of $10240^3$, and illustrate the advantages of feature-based meshing by running FEM simulations on its volumetric interior.


\color{black}
\noindent In summary, we make the following technical contributions:

\begin{itemize}[leftmargin=*, itemsep=3pt, topsep=3pt, parsep=0pt]
\item We introduce a novel algorithm for computing interior and exterior medial axis samples (which we term medial scaffolds) directly from a generic signed distance function (SDF), and demonstrate its performance to be on par with or better than state-of-the-art mesh-based medial axis computation techniques.
\item We introduce FOMOS, an LFS-based adaptive SDF meshing technique that reduces the time and memory complexity of discrete SDF meshing from cubic to quadratic, achieves state-of-the-art performance, and guarantees a manifold output mesh homeomorphic to the target surface.
\item We extend our meshing framework to generate manifold meshes from huge truncated signed distance fields (TSDFs) and to create near-instantaneous SDF previews.
\end{itemize}
\color{black}

\section{Related Work}

The related works for feature-aware meshing of SDFs can be classified into two categories: Algorithms for computing the medial axis (as we and most works in literature prevalently use the medial axis for estimating LFS), 
and SDF meshing algorithms (algorithms for creating meshes from signed distance functions).
To our knowledge, our end-to-end algorithm is the first to take an SDF as input and directly mesh it based on the local feature size rather than a grid.


\paragraph{Medial Axis Approximation.} The main challenge in local feature sizing functions is its dual demand of geometric fidelity and computational efficiency. The foundations of works in this direction lie in either curvature- or shape-diameter-driven adaptations \cite{persson2006mesh,shapira2008consistent} or in local feature size estimation. As the local feature size (LFS) is defined as the distance to the closest point on the medial axis \cite{amenta1998crust}, most works in this direction rely on approximate medial axis computation, requiring global knowledge of the shape. Hence, in this section, we discuss various methods in the literature for computing the medial axis.

Due to its usefulness in shape analysis, early works in medial axis computation date back to the late 1960s \cite{blum1967transformation}. However, due to its highly unstable nature and high geometric complexity, later works (in our context) explored the connection between a subset of easily computable Voronoi vertices and the medial axis \cite{amenta2001power,boissonnat2001natural}. Though theoretically valid when the boundary samples approach infinity, it cannot produce a clean medial axis from real-world inputs for practical usage. This led to several works focusing on pruning the extraneous branches from the medial axis \cite{amenta2001medial,dey2004approximating,miklos2010discrete,yan2018voxel}, \textcolor{black}{employing restricted power diagram \cite{wang2022computing,wang2024mattopo}} and simplifying the medial axis \cite{li2015q,huang2024dynamic,sun2015medial,guo2024medial,wang2026structural}, but these usually require user-defined parameters. Another line of work focuses on computing medial axis points by shrinking spheres \cite{ma20123d,jalba2012surface,dou2022coverage}, but assumes the closest points on the surface are known and well-defined. Recent methods utilize deep learning to predict the medial axis \cite{rebain2019lsmat,lin2021point2skeleton,11048747,kong2026quasi}. Despite these advances, medial axis computation remains costly and non-robust, making it unsuitable for our needs. In addition, unlike typical point cloud and mesh representations, the SDF representation we are handling poses additional challenges as it lacks explicit topology and exhibits gradient sensitivity near sharp features and in under-resolved regions. It is also worth noting that, although not directly related to or applicable to our objective, time-consuming LFS computations on point clouds \cite{fu2024lfs} and 2D Bézier curves \cite{ohrhallinger2023feature} have recently been studied in the literature. \textcolor{black}{A closely related method \cite{persson2006mesh} detects the medial axis as the shock set of an SDF approximated by polynomials along grid-cell edges; however, its cubic computational complexity limits its suitability for the fast, high-resolution meshing targeted in our work.}

 Unlike these methods, we prioritize computing a {\em medial scaffold} as a set of estimated medial centers over precise medial axis computation, focusing solely on meshing an SDF. These medial scaffolds are then used to estimate a more precise LFS. 
 Furthermore, the medial scaffolds can be computed arbitrarily sparsely, while still precisely meshing using the entire SDF, resulting in an output-sensitive mesh generation time, enabling instant previews with fewer triangles.

\paragraph{SDF Meshing Algorithms. }

While widely used in geometry processing, SDFs are often converted to meshes to facilitate further processing. The classical approach, Marching Cubes \cite{mc, chernyaev1995marching}, grids the SDF and applies a look-up table to generate triangulated surfaces following a predefined set of rules. This is further improved in various directions, for instance, to resolve ambiguities \cite{lorensen1998marching}, capture sharp features \cite{kobbelt2001feature,ju2002dual}, and to enhance memory efficiency \cite{vega2019fast}. However, these techniques often result in non-manifold surfaces and poor-quality triangles, limiting their utility in many further applications (see Figure \ref{fig:teaser}). 
Recent deep learning variants, Neural Marching Cubes \cite{chen2021neural} and Neural Dual Contouring \cite{chen2022neural}, being the most prominent among them, learn the tessellation templates and vertex positions to reconstruct higher-quality meshes with improved feature preservation. 
In this line of work, Deep Marching Cubes \cite{liao2018deep}, MeshSDF \cite{remelli2020meshsdf}, and FlexiCubes \cite{shen2023flexible} introduced techniques to offer differentiable isosurfacing procedures. \textcolor{black}{Although not their primary focus, some learning-based methods reconstruct surfaces from point clouds using an SDF as an intermediate representation \cite{zhang2025high,ren2023geoudf,zhou2024cap}. However, their generalization to complex, previously unseen geometries may be limited (as illustrated in Figure \ref{fig:teaser}), they are often time-consuming, and they often result in low-quality triangles.} 

Despite the surge in deep learning-based techniques, geometric algorithms remain popular due to their simplicity and generalizability; a comprehensive literature review is available in \cite{de2015survey}. 
A more recent work uses a shrinkwrap or inflation method that iteratively deforms an initial mesh towards the desired surface \cite{hanocka2020point2mesh,bukenberger2021water}.
A global approach, Reach for the Spheres (RFTS) \cite{sellan2023reach}, shrink-wraps an initial mesh around the SDF to fit the surface using a gradient flow (note that topology changes are not allowed during the process).
Its successor, Reach for the Arcs (RFTA) \cite{sellan2024reach}, manages to handle non-zero genus objects and is somewhat faster, but reconstructs many bad triangles. \textcolor{black}{Other approaches focus on adaptive sampling to reduce the cost of implicit-surface extraction. McGrids \cite{ren2024mcgrids} uses iterative Monte Carlo sampling to reduce implicit-field queries, but incurs additional grid-construction overhead, produces bad quality triangles, and may require denser sampling to recover sharp features than methods that exploit normal information. Similarly, power-diagram-based extraction \cite{wang2026power} adaptively refines the extracted surface according to approximation error, but does not explicitly incorporate local feature size into its refinement criterion or directly optimize triangle quality. Another related approach } 
\cite{kohlbrenner2025polyhedral} follows a similar direction, but computes the largest empty sphere around every voxel using polyhedral approximations. As shown in Figure \ref{fig:teaser}, although these techniques work very well, they are particularly designed for sparse inputs (e.g., $30^3$) and require a significant amount of time and/or memory for large inputs, rendering them unsuitable for interactive applications such as RibbonSculpt \cite{Sureshkumar2025RibbonSculpt}, which is the focus of our work. In addition, the results of existing methods are not specifically designed for feature-preserving sampling and often produce a large number of (mostly redundant) triangles. \textcolor{black}{Furthermore, they are designed specifically for SDFs represented on discrete grids, while ours can also handle continuous SDFs (for which we sample the boundary discretely, e.g., but not necessarily, on a grid).}

\section{Method}
\label{sec:method}
The main technical challenge we are addressing is the LFS computation from SDFs.
We then use it with a meshing algorithm to generate the desired surface/volume meshes from the SDFs. Thanks to the inherent property of feature-sized sampling constructed using the LFS, the resulting meshes are sampled proportionally to this local feature sizing function.
In this section, after giving some definitions, we describe how we create a mesh for the zero surface of the SDF with feature-proportional triangle sizes.
Then we show how to extend our method to a light-weight variant that handles huge inputs.

\subsection{Background and Definitions}
\label{sec:definitions}

First, we recall a few basic definitions in the continuous case to facilitate further understanding. Let the boundary $\Omega$ be a set of simple closed smooth (twice differentiable) 2-manifolds in $\mathbb{R}^3$,

\begin{enumerate}
    \item The medial axis $M$ of $\Omega$ is defined as the set of all points having at least two closest points in $\Omega$ \cite{blum1967transformation}
    \item The local feature size $\mbox{lfs}(p)$ at a point $p \in \Omega$ is the Euclidean distance from $p$ to the closest point in $M$ \cite{ruppert1993new}
    \item The union of the centers of all medial balls of $\Omega$ also forms the medial axis $M$ of $\Omega$, and their radii are an upper bound of the local feature size at the points where they touch $\Omega$.
    \item Given a closed boundary, medial balls $B(c,r)$ can be classified as non-overlapping {\em interior} and {\em exterior} balls. 
    \item The surface $\Omega$ is called $\varepsilon$-sampled by a point set $S$ if for each point $p \in \Omega$, $\exists s \in S\textbackslash p$ such that $|p,s|<\varepsilon\cdot\mbox{lfs}(p)$, with the $|\cdot,\cdot|$ operator denoting the Euclidean distance, for $\varepsilon \in (0,1]$.
\end{enumerate}

In our case, we define $H$ as a $0$-isosurface ($\phi(x) =0$) of a signed distance function (SDF) $\phi: \mathbb{R}^3 \to \mathbb{R}$, with $\phi(x)$ as $|x,H|$ with interior negative, and our medial scaffold $\hat{M}$ as set of medial centers.

\noindent \paragraph{Overview of our method.} Our method takes an SDF as input and generates a mesh satisfying the $\varepsilon$-sampling condition. We adopt this meshing algorithm  \cite{boissonnat2005provably}, customizing it to incorporate our novel LFS computation from the SDFs. Our method consists of three main steps: (1) Computing the medial scaffold that is required for (2) computing the LFS, which is used in turn for (3) LFS-based meshing. We first elaborate on our method in theory, then give implementation details, and finally, analyze its accuracy.

\subsection{Our Algorithm for the Continuous Case}

Since we assume $H$ to be smooth, we can fit maximally inscribed balls $B(c,r)$ such that they touch (tangentially) at least 2 points in $H$, without $r$ vanishing.
Since such a touching point $p$ is tangent to $H$, we can determine a $B(c,r)$ tangent to $H$ at $p$ by varying $r$ until it fulfills the criteria both of being empty of $H$ and touching at least one other point $q \in H$.
We exploit the fact that $\phi(c)$ directly gives us the (signed) distance to the closest point $x$ on $H$, with $x=c-\phi(c)\nabla\phi(c)$.
We start with a diameter for $B$ equal to the bounding box diameter of $H$ and bisect $r$ until $B$ touches $q$. While bisecting, we decrease its upper bound if $\phi(c)<r$ and else increase the lower bound for $r$ (then, $x=p$).
This guarantees that $B$ will converge to a medial ball.
\textcolor{black}{It is worth noting that a direct application of sphere shrinking~\cite{ma20123d} is not suitable in our setting, as it requires known closest points to fit the spheres and iteratively shrink them. In contrast, we obtain only the closest distance from the SDF, while projecting candidate medial points onto $H$ using the SDF gradient is prone to inaccuracies due to grid discretization. Our bisection procedure addresses this limitation and is guaranteed to converge in the limit of infinite grid resolution, independently of the sampling density.}

After all medial balls have been computed, the union of their centers forms the medial axis $M$, and we can compute $\mbox{lfs}(p)=|p,m|$ for $p \in H$ to its closest point $m \in M$.
Then, an $\varepsilon$-sampled point set $S \in H$ can be generated by adding samples $s \in H$ to $S$ near points $p \in H$ while $\exists p: |p,s| \geq \varepsilon\cdot\mbox{lfs}(p)$.

However, since we cannot test an arbitrary number of medial balls or samples, we describe a practical implementation below that, for the sake of simplicity, takes an SDF defined on a uniform grid as input, without loss of generality. \textcolor{black}{Our algorithm begins by computing a discrete set of sample points on the medial axis, which we term the \textit{medial scaffold} and denote by $\hat{M}$.}


\subsection{Computing the medial scaffold}


\begin{wrapfigure}{R}{\dimexpr2.1in - .95\columnsep\relax}
\vspace{-\intextsep}
\centering
\hspace*{-.95\columnsep}\includegraphics[width=1.6in]{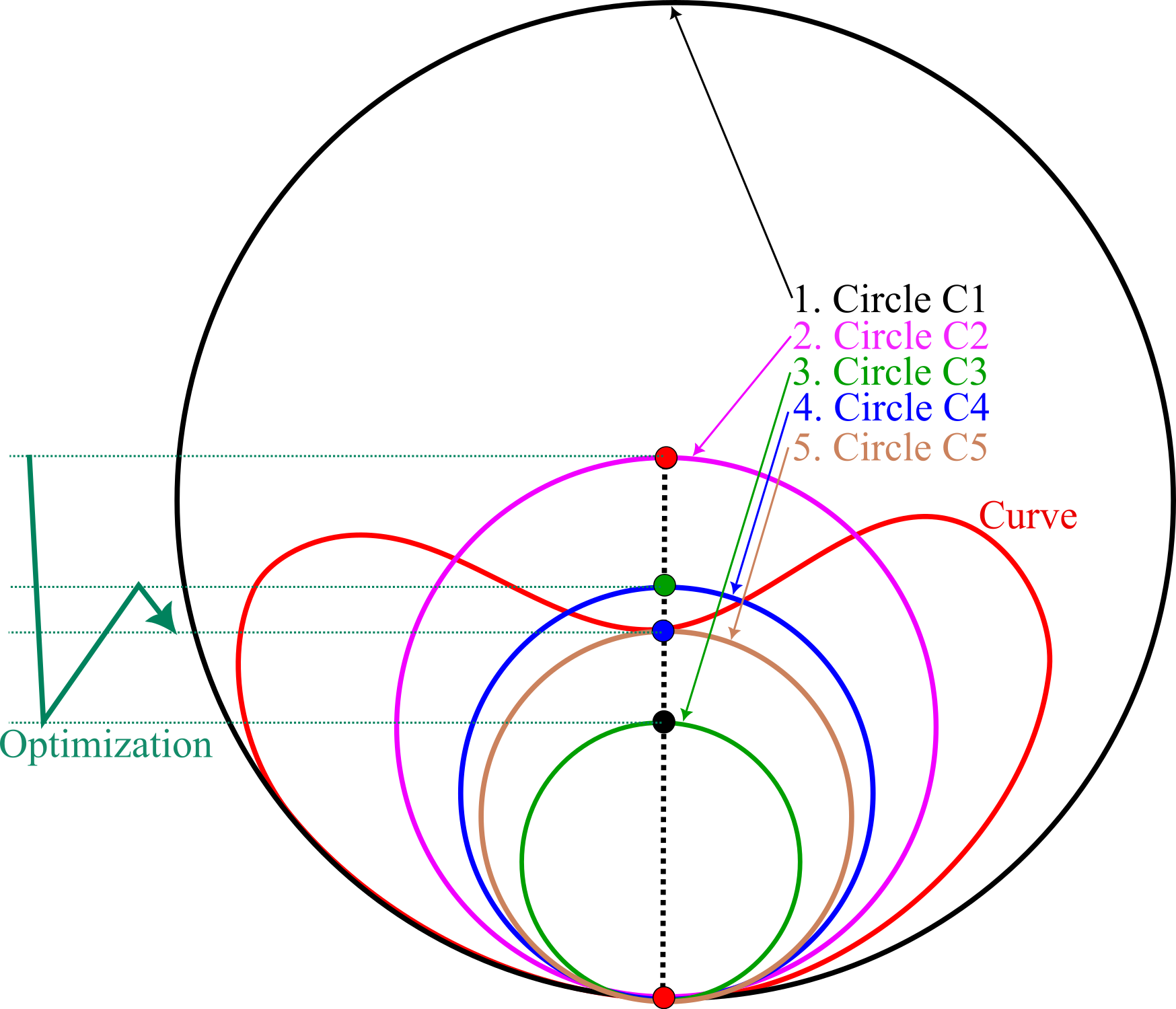}
\vspace{-10pt}
\end{wrapfigure}

The core contribution of FOMOS is the efficient computation of medial axis points directly from the zero surface of an SDF, rather than its volume, reducing both time and space complexity from cubic to quadratic (except for filtering initial footpoints).
Note that for the gridded SDF, $H$ is $C^2$-continuous everywhere except at cell boundaries ($C^0$).
First, we compute a dense, uniformly distributed set of discrete footpoint samples on $H$. For that, we only consider SDF grid values near the surface ($\phi$ smaller than the cell diameter, i.e., only at the corners of cells that contain the surface). These grid points are then projected onto $H$ using gradient descent, iteratively refining the position of each point by computing the gradient $\nabla \phi(p^k)$, and then moving towards $p^{k+1}=p^k-\phi(p^k)\nabla \phi(p^k)$. This directed movement rapidly (on average, 2.5-3.5 iterations) converges to near the surface, $\phi(p)<\mu$, with a precision of $\mu=10^{-2}d$, where $d$ is the grid cell diagonal $d=\sqrt{3}|e|$ for a cell edge $e$.
The projected footpoints lie at most at a distance $d$ from their grid corner, yielding a maximum spacing of $2d+|e|$ in Euclidean space, but in practice their RMS distance is $<d/3$, so the footpoint density is quite uniform on $H$.

\color{black}
\begin{algorithm}[t]
\caption{\textcolor{black}{FOMOS: LFS computation from an SDF}}
\label{alg:fomos}
\begin{algorithmic}[1]
\color{black}
\Require SDF $\phi$ to zero-set boundary $H$, a uniform grid with cell edge $e$; tolerance $\mu = 10^{-2}d$, $d=\sqrt{3}|e|$
\Ensure Medial scaffold $\hat{M}$, LFS estimate $\hat{\mbox{lfs}}(\cdot)$
\Statex \textit{// (1) Footpoint sampling on $H$}
\State $P \gets \{\, g \in \text{grid corners} \mid |\phi(g)| < d \,\}$
\ForAll{$p \in P$}
    \Repeat \;$p \gets p - \phi(p)\,\nabla\phi(p)$ \Comment{$\approx$2.5--3.5 iterations}
    \Until{$|\phi(p)| < \mu$}
\EndFor
\Statex \textit{// (2) Medial scaffold via tangent-ball bisection}
\State $\hat{M} \gets \emptyset$
\ForAll{$p \in P$}
    \State $n \gets \nabla\phi(p)$
    \State $(c_{in}, r_{in}) \gets \Call{FitBall}{p, -n}$;
    \State $(c_{out}, r_{out}) \gets \Call{FitBall}{p, +n}$
    \State $\hat{M} \gets \hat{M} \cup \{\, \arg\min_{c \in \{c_{in}, c_{out}\}} r \,\}$ \Comment{keep smaller ball}
\EndFor
\Statex \textit{// (3) Local feature size}
\State $\mathcal{T} \gets \Call{KdTree}{\hat{M}}$
\State $\hat{\mbox{lfs}}(x) := |x, \mathcal{T}.nn(x)|$ \Comment{$O(\log|\hat{M}|)$ per query}
\State \Return $\hat{M}, \hat{\mbox{lfs}}$
\Statex
\Function{FitBall}{$p, v$}
    \State $r_{lo} \gets 0$;\; $r_{hi} \gets \tfrac{1}{2}\,\mathrm{diag}(BB(H))$
    \Repeat
        \State $r \gets (r_{lo}+r_{hi})/2$;\; $c \gets p + r\,v$
        \State $q \gets c - \phi(c)\,\nabla\phi(c)$ \Comment{closest point on $H$}
        \If{$|\phi(c)| < r$ \textbf{and} $|p,q| \geq d$}
            \State $r_{hi} \gets r$ \Comment{ball hits $H$ elsewhere: shrink}
        \Else
            \State $r_{lo} \gets r$ \Comment{empty, or $q \approx p$: grow}
        \EndIf
    \Until{$\big|r - |\phi(c)|\big| < \mu$ \textbf{or} $r_{hi}-r_{lo} < \mu$} \Comment{$\approx$6--10 iterations}
    \State \Return $(c, r)$
\EndFunction
\end{algorithmic}
\end{algorithm}
\color{black}

\begin{figure*}
    \includegraphics[width=\linewidth]{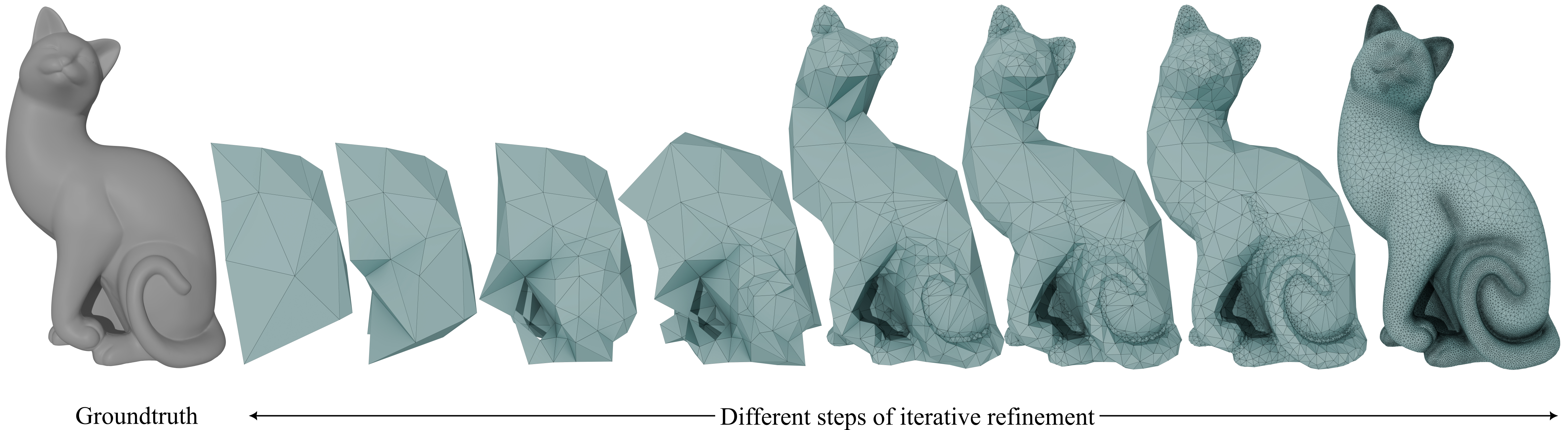}
    \caption{Evolution of the meshing from a few triangles based on vertices on the SDF zero surface until the $\varepsilon$-sampling criterion $\varepsilon=0.2$ is fulfilled.}
      \label{fig:evolution}
\end{figure*}

Once the position of $p$ on the surface is determined, we compute the largest empty tangent balls both in its gradient direction and the opposite. 
The sphere fitting for computing the LFS, as illustrated in the inset, has these steps:

\begin{itemize}
    \item{Initialization:} A tangent ball $B(c,r)$ with $r=\frac{1}{2}diag(BB(H))$
    \item{Radius search:} We iteratively bisect $[0,r]$ until $|r-\phi(c)|<\mu$ or the bounds converge to $\mu$ ($\varnothing$ 6-10 iterations, $16^3-1024^3$)
    \item{Near footpoint $p$:} If $c$'s closest point on $H$, $q=c-\phi(c)\nabla\phi(c)$, has $|p,q|<d$, it can be in the same grid cell, and we reject it due to trilinear interpolation artifacts of $\phi$ inside cells plus overproportional sphere fitting errors occurring near $p$
    \item{Radius assignment:} We pick the smaller of these two balls
\end{itemize}

By precomputing gradients on the SDF grid, we cut runtimes roughly in half while increasing memory usage to 4x. This can, however, be switched off in our method for low-memory machines.
Note that while a gridded SDF is $C^0$ at cell boundaries, our medial axis is very accurate, as we will analyze in detail in Section \ref{sec:accuracy}.


\subsection{Computing the LFS} As defined above, the local feature size at a point $p$ on a boundary $\Omega$ is $\mbox{lfs}(p)=|p,m|$ with $m \in M$ as the closest point to $p$ in the medial axis $M$ of $\Omega$.
For our medial scaffold $\hat{M}$ as an approximate sample of $M$, we construct a kd-tree to efficiently query the nearest neighbor $nn(p)$ for any point $p \in \mathbb{R}^3$, in $O(log |\hat{M}|)$ time complexity. Using this, we can define our LFS as $\hat{\mbox{lfs}}(p)=|p,nn(p)|$ (see Algorithm \ref{alg:fomos}).

\subsection{Meshing the SDF with Local Feature Size (LFS)}

\begin{figure*}
    \includegraphics[width=\linewidth]{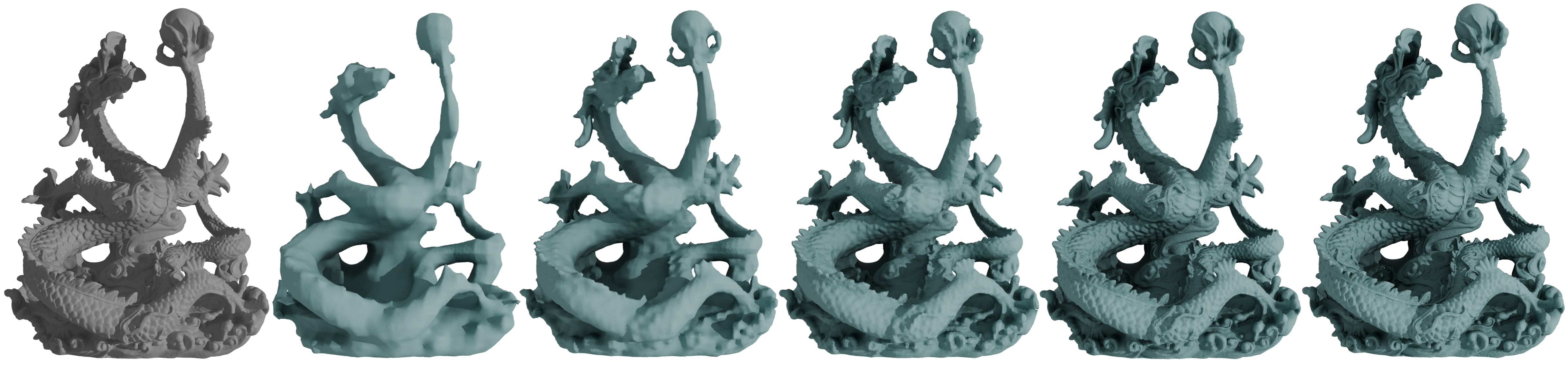}
    \caption{FOMOS on $32^3$, $64^3$, $128^3$, $256^3$, and $512^3$ inputs of a detailed model (left).}
      \label{fig:resolutions}
\end{figure*}

\textcolor{black}{Our meshing algorithm builds on Boissonnat and Oudot~\cite{boissonnat2005provably}, which was in turn inspired by Chew~\cite{chew1993guaranteed}. The input is a closed boundary $\Omega$ accessed only through an oracle (function) that reports whether a point lies inside or outside, and the objective is to create a surface mesh that preserves the geometry and topology of $\Omega$. Their algorithm initializes a three-dimensional Delaunay triangulation $Q$ from a sparse set of points on $\Omega$. Using an oracle, it identifies intersections between $\Omega$ and Voronoi edges that are dual to the triangles of $Q$. The surface approximation is then defined as the restricted Delaunay triangulation $T_R \subseteq Q$, comprising precisely those triangles whose dual Voronoi edges intersect $\Omega$. The algorithm iteratively refines $T_R$: when a triangle $t$ violates a refinement criterion (e.g., a minimum edge-length constraint), the corresponding intersection point $x$ is inserted into $Q$, and edge flips are performed as necessary to maintain the Delaunay property. Boissonnat and Oudot also considered $\varepsilon$-sampling as a refinement criterion, and demonstrated exact LFS evaluation for simple algebraic surfaces. For more general shapes, they proposed using level-set thinning~\cite{chen1993parallel} or the $\lambda$-medial axis~\cite{chazal2005lambda}, which do not guarantee sufficiently accurate local feature-size estimates for general shapes, potentially leading to inappropriate sampling densities and compromising the guarantees of $\varepsilon$-sampling.}

\textcolor{black}{We use our LFS estimator (Section~\ref{sec:method}) to apply the $\varepsilon$-sampling criterion directly to SDF representations. Given a Delaunay triangulation $Q$, an oracle finds the intersection $p = e \cap H$ between each Voronoi edge $e$ and the zero level set $H$ of the SDF.}

\color{black}
For each triangle $t$ in the restricted Delaunay triangulation, we evaluate the local sampling ratio:

\begin{equation}
   \varepsilon_t = \frac{r_t}{\hat{\operatorname{lfs}}(p)} 
\end{equation}

where $r_t$ denotes the radius of the associated Voronoi structure and $\hat{\operatorname{lfs}}(p)$ is our estimated local feature size at $p$. This ratio measures the sampling scale relative to the local geometric feature size: larger values indicate that the surface may be insufficiently sampled in that region. Whenever $\varepsilon_t$ exceeds the prescribed threshold $\varepsilon$, we insert $p$ into $Q$, splitting $t$ and locally refining the mesh. This process continues until the sampling criterion is satisfied. Figure~\ref{fig:evolution} illustrates the refinement process for $\varepsilon = 0.2$, while Figure~\ref{fig:resolutions} shows the results obtained at different grid resolutions.
\color{black}

\subsection{Sampling Guarantee}
\label{sec:guarantee}
\color{black}
FOMOS only tests the $\varepsilon$-sampling criterion at the intersection points of the restricted Delaunay (surface) triangles with their Voronoi edges, which are inserted as refinement vertices if they do not meet the sampling condition.
We now show that the mesh produced by our FOMOS satisfies the $\varepsilon$-sampling criterion on all of $H$, not only at these tested points.

\paragraph{Notation.}
Let $S$ be the current vertex set and $\delta(x) = \min_{s \in S} |x,s|$ the distance
from $x$ to the nearest vertex. For a restricted Delaunay triangle
$t = (v_1,v_2,v_3) \in T_R$ with dual Voronoi edge $e$, let $p_t = e \cap H$ be its
refinement point and $r_t = |p_t, v_i|$ the radius of its surface Delaunay ball
$B(p_t, r_t)$. Let $R_t$ and $c_t$ denote the circumradius and circumcenter of $t$.
Refinement inserts $p_t$ whenever $r_t \ge \varepsilon\,\hat{\mbox{lfs}}(p_t)$.
We assume our LFS estimate has (small) bounded relative error,
\begin{equation}
  \mbox{lfs}(x) \;\le\; \hat{\mbox{lfs}}(x) \;\le\; (1+\alpha)\,\mbox{lfs}(x),
  \label{eq:lfs-accuracy}
\end{equation}
where the lower bound holds because $\hat{M}$ is a subset of samples of $M$
(see Section~\ref{sec:accuracy} for an estimation of $\alpha$). We write
$\tilde\varepsilon = (1+\alpha)\,\varepsilon$.

\paragraph{Properties of the local feature size}

First, we show that the LFS varies slowly across the surface, never more than the distance:

\begin{lemma}[Lipschitz continuity~{\cite[Lemma~1]{amenta1998surface}}]
\label{lem:lipschitz}
For all $p, q \in H$: $\;\mbox{lfs}(p) \le \mbox{lfs}(q) + |p,q|$.
\end{lemma}

\noindent Then, we show that the surface is locally flat for small $\varepsilon$:

\begin{lemma}[Normal variation~{\cite[Lemma~2]{amenta1998surface}}]
\label{lem:normals}
For $p, q \in H$ with $|p,q| \le \rho\,\mbox{lfs}(p)$ and $\rho < 1/3$, the angle
between the surface normals at $p$ and $q$ is at most $\rho/(1-3\rho)$.
\end{lemma}

\paragraph{Sampling at refinement points}

The third lemma specifies that each Voronoi edge intersection with its dual restricted Delaunay triangle fulfills our $\varepsilon$-sampling condition:


\begin{lemma}[Refinement points]
\label{lem:refpoint}
Upon termination, every $t \in T_R$ satisfies
$\delta(p_t) = r_t < \tilde\varepsilon\,\mbox{lfs}(p_t)$.
\end{lemma}

\begin{proof}
Since $p_t$ lies on the Voronoi edge dual to $t$, its nearest vertices in $S$ are
exactly $v_1, v_2, v_3$, all at distance $r_t$. Equivalently, $B(p_t, r_t)$ contains
no vertex of $S$ in its interior. Hence $\delta(p_t) = r_t$. Upon termination, the
refinement test holds for $t$, so by~\eqref{eq:lfs-accuracy},
$r_t < \varepsilon\,\hat{\mbox{lfs}}(p_t) \le \tilde\varepsilon\,\mbox{lfs}(p_t)$.
\end{proof}

\paragraph{Sampling inside triangles.}
The refinement point alone does not bound $\delta$ in its neighborhood. If $t$ is
obtuse, $c_t$ lies outside $t$, and $\delta$ increases when moving from $p_t$ away
from $t$. The following lemma shows that nonetheless every point within a triangle,
obtuse or not, is close to one of its vertices.

\begin{lemma}[Nearest-vertex distance in a triangle]
\label{lem:triangle}
For every $x \in t$: $\;\min_i |x,v_i| \le R_t \le r_t$.
\end{lemma}

\begin{proof}
Let $f(x) = \min_i |x,v_i|$, which is continuous on the compact set $t$, and
consider the Voronoi diagram of $\{v_1,v_2,v_3\}$ in the plane of $t$.
Within the Voronoi cell of a single $v_i$, $f(x) = |x,v_i|$ has no interior local
maximum. Along a bisector of $v_i v_j$, $f$ is convex, so its maximum over the
bisector segment within $t$ lies at an endpoint of that segment.
Hence the maximum of $f$ over $t$ is attained either (i) on an edge $v_i v_j$,
where $f \le |v_i v_j|/2 \le R_t$ because every edge is a chord of the
circumcircle, or (ii) at the Voronoi vertex $c_t$, if $c_t \in t$, where
$f = R_t$. Finally, $p_t$ lies on the line through $c_t$ orthogonal to $t$, so
$r_t^2 = R_t^2 + |p_t,c_t|^2$ and thus $R_t \le r_t$.
\end{proof}

\paragraph{Global guarantee.}
Let $\pi: \mathbb{R}^3 \to H$ denote the closest-point projection onto $H$, and define
\begin{equation}
  \kappa = \max_{t \in T_R}\, \max_{x \in t}\, \frac{|x,\pi(x)|}{r_t}.
  \label{eq:kappa}
\end{equation}

Now we are ready to prove that $\varepsilon$-sampling applies to each point of the entire surface $H$:

\begin{theorem}[Global $\varepsilon$-sampling]
\label{thm:global}
Assume that upon termination $\pi$ maps $T_R$ onto
$H$~{\cite[Theorem~4]{amenta1998surface}}, and that
$(1+\kappa)\,\tilde\varepsilon < 1$. Then for every $q \in H$,
\[
  \frac{\delta(q)}{\mbox{lfs}(q)} \;<\;
  \varepsilon' := \frac{(1+\kappa)\,\tilde\varepsilon}{1-(1+\kappa)\,\tilde\varepsilon},
\]
i.e., $H$ is $\varepsilon'$-sampled by $S$.
\end{theorem}

\begin{proof}
Let $q \in H$. Since $\pi$ is onto, there exist $t \in T_R$ and $x \in t$ with
$\pi(x) = q$, so $|x,q| \le \kappa\, r_t$ by~\eqref{eq:kappa}. Write $p = p_t$.

\emph{Distance to the nearest vertex.} By the triangle inequality and
Lemma~\ref{lem:triangle},
\[
  \delta(q) \le \min_i |q,v_i| \le \min_i |x,v_i| + |x,q|
  \le R_t + \kappa\, r_t \le (1+\kappa)\, r_t .
\]

\emph{Distance to the refinement point.} The vertices of $t$ lie on the boundary of
$B(p, r_t)$, so by convexity $t \subset B(p, r_t)$ and $|p,x| \le r_t$. Hence
$|p,q| \le |p,x| + |x,q| \le (1+\kappa)\, r_t$.

\emph{Local feature size.} By Lemma~\ref{lem:lipschitz} and Lemma~\ref{lem:refpoint},
\[
  \mbox{lfs}(q) \;\ge\; \mbox{lfs}(p) - |p,q|
  \;\ge\; \mbox{lfs}(p) - (1+\kappa)\, r_t
  \;>\; \big(1-(1+\kappa)\,\tilde\varepsilon\big)\,\mbox{lfs}(p).
\]

\emph{Combination.} Using Lemma~\ref{lem:refpoint} once more,
\[
  \delta(q) \le (1+\kappa)\, r_t
  < (1+\kappa)\,\tilde\varepsilon\,\mbox{lfs}(p)
  < \frac{(1+\kappa)\,\tilde\varepsilon}{1-(1+\kappa)\,\tilde\varepsilon}\,\mbox{lfs}(q).
\]
\end{proof}


Finally, we specify which bounds apply to each input value of $\tilde{\varepsilon}$:

\begin{corollary}[Bounds on $\varepsilon'$]
\label{cor:bounds}
Under the assumptions of Theorem~\ref{thm:global}:
\begin{enumerate}
  \item Unconditionally $\kappa \le 1$, hence
        $\varepsilon' \le 2\tilde\varepsilon/(1-2\tilde\varepsilon)$ for
        $\tilde\varepsilon < 1/2$.
  \item If $H$ is locally flat above each triangle, then $\kappa = O(\tilde\varepsilon)$
        and $\varepsilon' = \tilde\varepsilon/(1-\tilde\varepsilon) + O(\tilde\varepsilon^2)$.
\end{enumerate}
\end{corollary}

\begin{proof}
(1) For $x \in t$, $|x,\pi(x)| = \mbox{dist}(x,H) \le |x,p_t| \le r_t$, since
$p_t \in H$ and $t \subset B(p_t, r_t)$.
(2) For $\tilde\varepsilon < 1/3$, Lemma~\ref{lem:normals} applied within
$B(p_t, r_t)$ bounds the deviation of the normals of $H$ above $t$ by
$\tilde\varepsilon/(1-3\tilde\varepsilon)$. Hence $H$ is a graph over the plane
of $t$ there, with height $O(\tilde\varepsilon)\, r_t$ above
it~\cite{amenta1998surface}. Substituting $\kappa = O(\tilde\varepsilon)$ into
Theorem~\ref{thm:global} yields the claim.
\end{proof}

\noindent \emph{Remarks.}
Obtuse triangles require no special treatment. Points of $H$ beyond the long edge of
an obtuse $t$, where $\delta$ may exceed $r_t$, lie above an adjacent triangle $t'$.
Theorem~\ref{thm:global} bounds them via $r_{t'}$, which is itself tested.
For small $\varepsilon$, the guarantee $\varepsilon' \approx \tilde\varepsilon$ is
close to the prescribed threshold. In the limit $\kappa \to 0$, $\varepsilon' < 1$ holds
for $\tilde\varepsilon < 1/2$. Unconditionally (Corollary~\ref{cor:bounds}), it holds
for $\tilde\varepsilon < 1/3$.

\color{black}

Thanks to the intelligent meshing algorithm \cite{boissonnat2005provably} that is guaranteed to terminate\textcolor{black}{, as its subdivision stops either when our feature-sized criterion is fulfilled or at a minimum edge length proportional to grid cell size}, the resulting surfaces are, by definition, watertight, 2-manifold, and homeomorphic to the original surface. \textcolor{black}{Since our modification only changes the refinement criterion, the properties guaranteed by the underlying meshing algorithm are preserved. These properties are essential for many applications but are not guaranteed by most state-of-the-art methods.}

\subsection{Accuracy Analysis of our Method}
\label{sec:accuracy}

\begin{figure}[h]
    \centering
    \begin{adjustbox}{max width=\linewidth}
        \includegraphics[width=\linewidth]{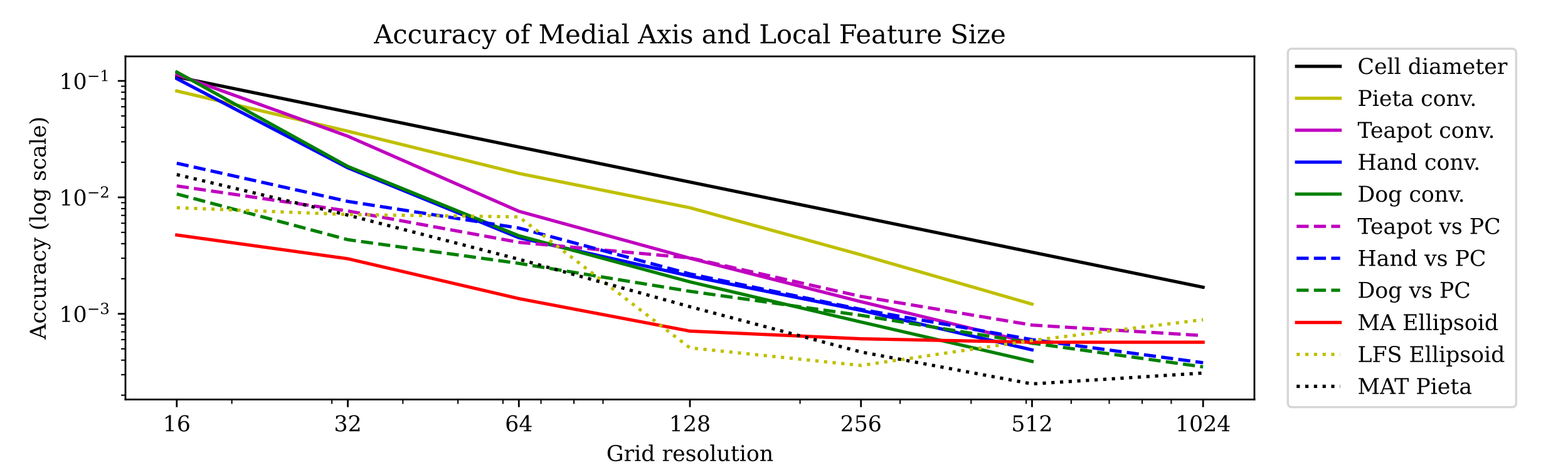}
    \end{adjustbox}
    \caption{Analysis of RMSE for: MA and LFS of {\scshape Ellipsoid} (analytical), MA convergence of four models to $1024^3$, convergence of our {\em interior} MA to {\scshape PowerCrust}'s for three models, and MAT--surface distance for {\scshape Pieta}.}
    \label{fig:accuracy}
\end{figure}



The accuracy of the medial axis, subsequently the LFS, and, further, $\varepsilon$-sampling, cannot be easily analyzed for non-analytical surfaces. Therefore, we first compute the error of both our medial axis and LFS using an analytical surface, then show how our medial axis both closely converges for increasing resolution of the SDF grid and towards a guaranteed homotopic surface, then compute LFS error using the medial axis transform, a by-product of our method, and finally, how the accuracy of $\varepsilon$ depends on the LFS (all in Figure \ref{fig:accuracy}). Note that the numerical errors arising from double-precision arithmetic are negligible compared with those introduced by grid discretization.

\begin{figure}
    \centering
    \includegraphics[width=\linewidth]{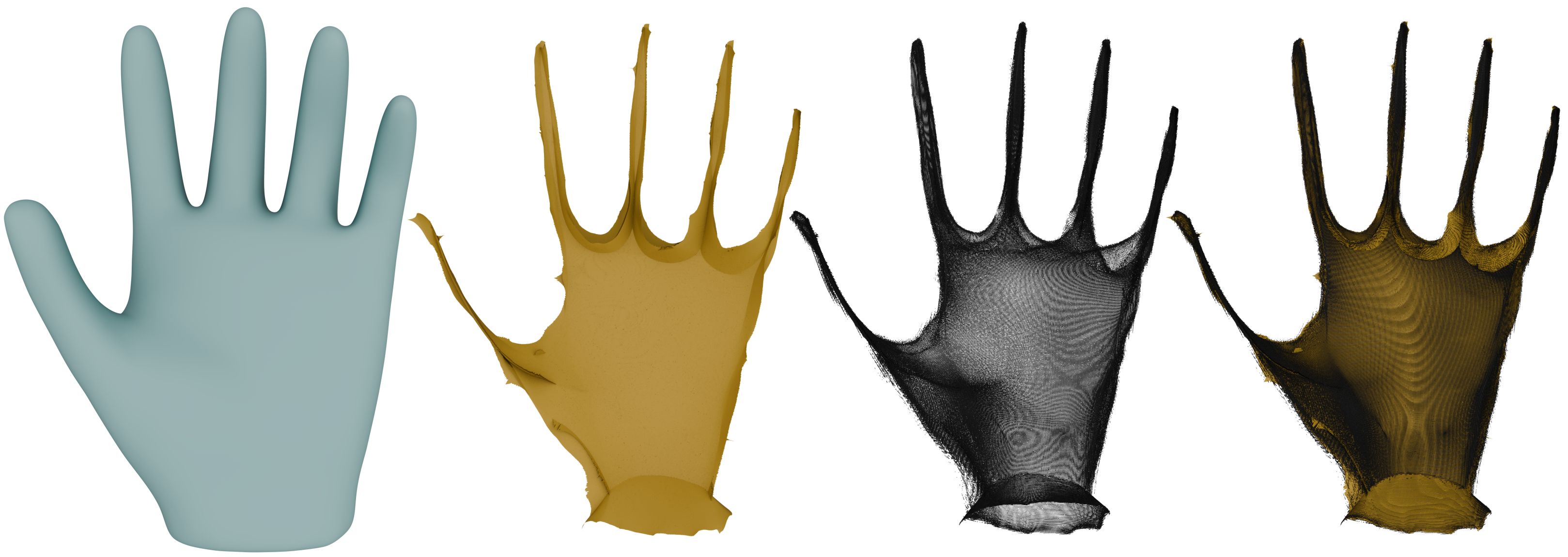}
    \caption{{\scshape Hand} model, from left to right: finely (6x) tessellated surface (2m vertices), {\scshape PowerCrust} and Our interior medial axes, and their overlap.}
    \label{CompPC}
\end{figure}


\noindent \paragraph{Analytical accuracy analysis of our medial axis}

To show the accuracy of the computation of our medial axis as well as its dependent local feature size, we analytically determine these values for an ellipsoid (a=0.9, b=0.7, c=0.5) centered at the origin, for which the medial axis is its evolute, an ellipse lying in the z-plane.
We intersect a line from its footpoint samples along their gradients with the z-plane to compute the medial axis ground truth.
For the local feature size ground truth, we project the footpoint onto the z-plane and compute its distance to the evolute if it is outside, or simply take $|z|$ if it is inside.
Both the error of the medial axis and the local feature size are always far below the grid cell diameter.

\noindent \paragraph{Convergence behavior of our medial axis}
To validate that our medial axis converges with increasing grid resolution, we compute the RMS distance between corner points on varying-sized grids to the same corner points on a baseline $1024^3$ grid, showing good convergence and error always below grid cell size.

\noindent \paragraph{Comparison with PowerCrust medial axis}
We also compare with the {\scshape PowerCrust} medial axis that converges to the actual medial axis (however, slow and only for extremely smooth sampled surfaces, making it unusable for the application we target). We Loop-subdivide three models to tessellation level 6, resulting in $\approx$1m triangles with average edge length $\approx$ cell diameter for $1024^3$.
The RMS Hausdorff distances of our interior medial axis to {\scshape PowerCrust} show converging behavior. See overlap for $1024^3$ in Figure \ref{CompPC}.

\noindent \paragraph{Medial axis transform analysis} We create an approximation of the shape by taking the surface $G$ of the union of all interior medial balls $\bigcup_{i=1}^n B_i(m_i, \phi(m_i))$ computed from medial scaffold points $m_i \in \hat{M}$.
Then, we sample $G$ by intersecting it with 1k line segments on random points of opposing faces of the bounding box for alternating axes.
The SDF values of these intersection points on $G$ yield their respective (Hausdorff) distances to $H$; we compute the RMSE between the medial axis transform (MAT) and the surface w.r.t. the bounding-box diagonal for {\scshape Pieta} to show its accuracy. 

\noindent \paragraph{$\varepsilon$-sampling analysis}
The error of $\varepsilon$ is dependent on the LFS error.
Since $|p,s|<\varepsilon\cdot\mbox{lfs}(p)$, the good accuracy of the LFS is indirectly proportional to $\varepsilon$.
Small LFS values at fine features that span just a few cells will proportionally increase their relative error, and with it, the error of the related $\varepsilon$, but for larger LFS values, with a medial ball radius $>8$, the $\varepsilon$ error is negligible with $0.1-1\%$, especially for the applications shown later, which do not require exact $\varepsilon$ values.
While we evaluate the SDF for a similar number of near-surface corner points as Marching Cubes, the mesher queries fewer vertices (187k vs 16k for {\scshape Buddha} $128^3$ with $\varepsilon=0.3$ in the teaser).

\color{black}
\noindent \paragraph{Comparison with mesh-based medial axis computation techniques.} Table~\ref{tab:reconstruction_results} compares surface reconstruction quality from medial representations computed by FOMOS and four mesh-based medial axis methods: MATFP \cite{wang2022computing}, VMAS \cite{huang2024dynamic}, MATTopo \cite{wang2024mattopo}, and Structural MAT \cite{wang2026structural}. We evaluate reconstruction quality using Chamfer Distance (CD), root mean square error (RMSE), F-Score, and normal consistency (NC). Reconstruction accuracy provides an indirect measure of how well each computed medial representation preserves the geometric information required to recover the original surface. To make the comparison with mesh-based approaches more consistent, we convert the input meshes to signed distance fields (SDFs) and restrict our medial representation to the interior medial scaffold for this experiment. At a resolution of \(512^3\), FOMOS achieves the lowest CD (\(7.5\times10^{-4}\)) and RMSE (\(0.00104\)) among the evaluated methods, improving upon Structural MAT by approximately 20.2\% and 14.8\%, respectively. FOMOS also achieves the highest NC (98.16) and a near-perfect F-Score (99.99), demonstrating that its computed medial scaffold retains sufficient geometric information for highly accurate surface reconstruction. Its reconstruction quality remains competitive at \(256^3\), whereas coarser resolutions lead to increasingly pronounced errors, particularly at \(64^3\). Overall, these results show that the medial representation computed by FOMOS supports surface reconstruction at a quality comparable to or better than that of the evaluated medial-axis methods, depending on the metric and resolution.

The comparison involves methods that differ in both their input representations and their construction of medial structures. Existing approaches are evaluated using mesh-based representations, whereas FOMOS operates on SDFs. We therefore convert the input meshes to SDFs and extract only the interior medial scaffold for this experiment, rather than using the full representation produced by FOMOS. This provides a more comparable evaluation of interior-based surface reconstruction while preserving the distinction between the underlying pipelines. In particular, FOMOS constructs a medial scaffold rather than computing a medial axis alone, and its full formulation represents both interior and exterior medial structures. Accordingly, the results should be interpreted as a comparison of the geometric fidelity of surfaces reconstructed from the respective medial representations, rather than as evidence that the methods compute equivalent representations or follow identical procedures.

\begin{table}[htb]
\centering
\resizebox{0.8\linewidth}{!}{%
\begin{tabular}{lrrrr}
    \toprule
    \textbf{Method}  &
    \textbf{CD} ($\downarrow$) &
        \textbf{RMSE} ($\downarrow$) &
    \textbf{F-Score} ($\uparrow$) &
    \textbf{NC} ($\uparrow$) \\
    \midrule
    \rowcolor{lightgray}MATFP & $9.1\times10^{-4}$ & 0.00156& \underline{99.98} & 97.98 \\
    VMAS & $1.08\times10^{-3}$ &0.00135 & 99.53  & 97.63  \\
    \rowcolor{lightgray} MAT Topo & $1.15\times 10^{-3}$ & 0.00162 & 99.08& 96.92\\
    Structural MAT & \underline{$7.29\times10^{-4}$} & \underline{0.00095} & \textbf{99.99} & \underline{98.15}\\
    
        \midrule \rowcolor{lightgray}FOMOS ($512^3$) & $\boldsymbol{6.1\times10^{-4}}$ & \textbf{0.00082}  & \textbf{99.99} & \textbf{98.16}\\
    FOMOS ($256^3$)  & $8.12\times10^{-4}$ & 0.00105 & 99.97 & 98.05\\
    \rowcolor{lightgray}FOMOS ($128^3$) & $1.61\times10^{-3}$ & 0.0023 & 99.44 & 97.32 \\
    FOMOS ($64^3$) & $1.72\times10^{-2}$ & 0.03413& 76.83  & 87.41\\
    \bottomrule
\end{tabular}%
}
\caption{\textcolor{black}{Quantitative evaluation of surface reconstructions from medial axes and radial functions using Chamfer Distance (CD), Root Mean Square Error (RMSE), F-Score, and Normal Consistency (NC). Best results are shown in bold and second-best results are underlined.}}
\label{tab:reconstruction_results}
\end{table}

\begin{figure}
    \centering
    \includegraphics[width=0.8\linewidth]{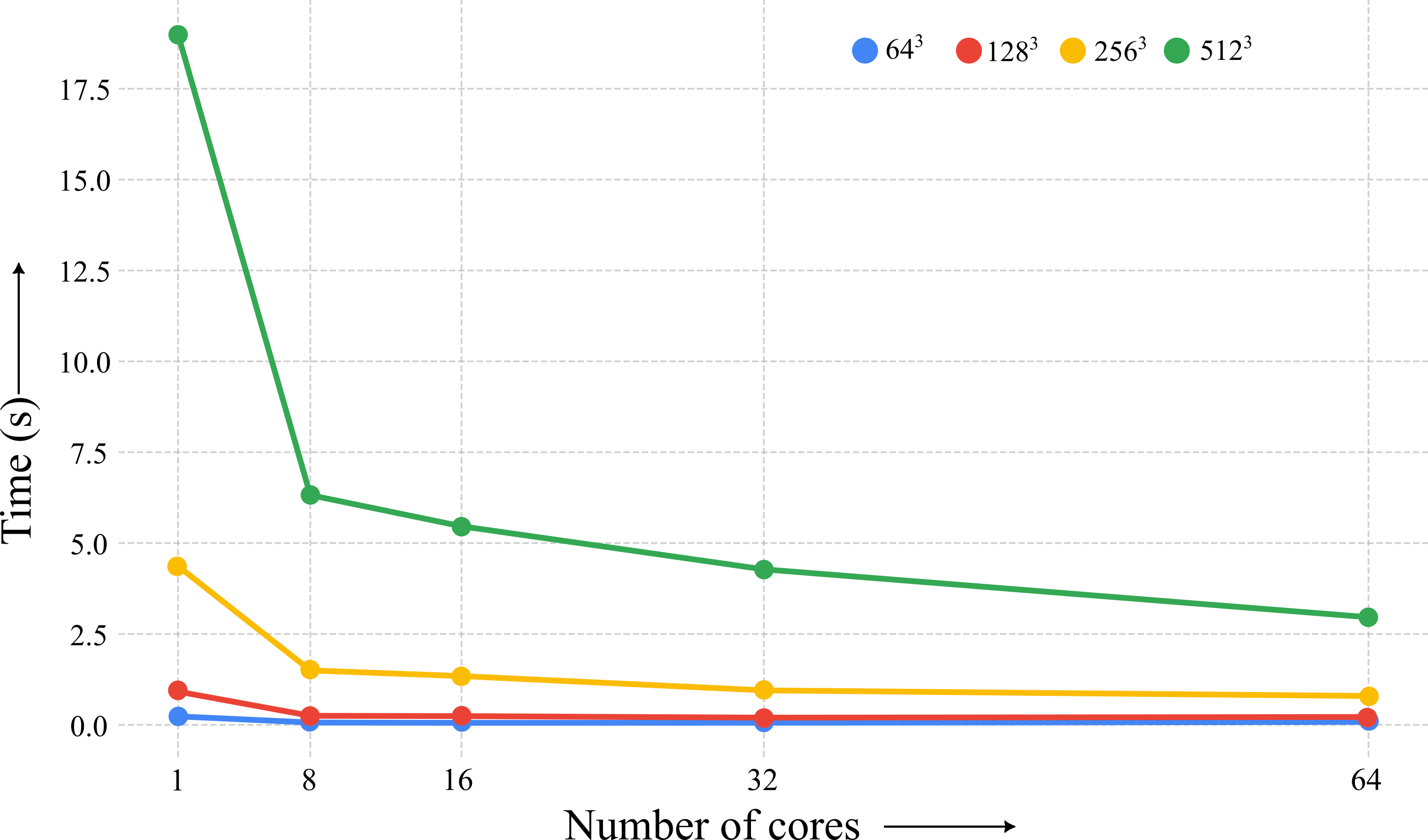}
\caption{\textcolor{black}{Runtime of medial scaffold computation for different grid resolutions and numbers of CPU cores.}}
    \label{fig:threads}
\end{figure}

\paragraph{Scalability of Medial Scaffold Computation: } Figure \ref{fig:threads} demonstrates the effect of multithreading on medial scaffold computation across different grid resolutions. While smaller grids exhibit diminishing returns after 8-16 cores, larger workloads continue to benefit from increased parallelism. For the $512^3$ grid, the execution time decreases from 18.99s to 2.96 on 64 cores, corresponding to a $6.41\times$ speedup and an $84.4\%$ reduction in runtime. Similarly, the $256^3$ grid achieves a $5.50\times$ speedup on 64 cores. In contrast, $64^3$ and $128^3$ grids attain their best measured performance at 16 and 32 cores, respectively, suggesting that parallelization overhead becomes increasingly significant relative to the computational workload at lower resolutions. These results demonstrate the scalability of the medial scaffold computation for high-resolution grids and indicate that the number of threads should be selected based on the workload size.

\color{black}

\subsection{Handling Huge Sparse Inputs}
\label{sec:huge}

\begin{table}[htb]
    \centering
    \resizebox{\linewidth}{!}{
    \begin{tabular}{ lrrrrrrr } 
        \toprule
        \textbf{Model} & \textbf{k} & \textbf{Medial (s)} & \textbf{Mesh (s)} & \textbf{Total (s)} & \textbf{Memory} & \textbf{\# Medial} & \textbf{\# Tri} \\
        \bottomrule
        \rowcolor{lightgray}
{\scshape EMU} & 32 & 0.02 & 0.88 & 0.90 & 1345m & 0.4k & 12k \\
{\scshape EMU} & 16 & 0.12 & 2.14 & 2.26 & 1345m & 3k & 32k \\
        \rowcolor{lightgray}
{\scshape EMU} & 8 & 1.01 & 5.82 & 6.83 & 1347m & 20k & 92k \\
{\scshape EMU} & 4 & 6.57 & 14.05 & 20.62 & 1598m & 146k & 261k \\
        \rowcolor{lightgray}
{\scshape EMU} & 2 & 66.29 & 62.13 & 128.42 & 3394m & 1162k & 1093k \\
{\scshape EMU} & 1 & 960.46 & 332.37 & 1292.83 & 14053m & 9347k & 5272k \\
        \bottomrule
        \rowcolor{lightgray}
{\scshape Thai} & 32 & 0.21 & 3.50 & 3.71 & 14828m & 2k & 37k \\
{\scshape Thai} & 16 & 1.86 & 12.56 & 13.42 & 14832m & 19k & 132k \\
        \rowcolor{lightgray}
{\scshape Thai} & 8 & 10.72 & 25.56 & 36.28 & 14854m & 157k & 399k \\
{\scshape Thai} & 4 & 132.03 & 78.22 & 201.25 & 15039m & 1258k & 1006k \\
        \rowcolor{lightgray}
{\scshape Thai} & 2 & 2654.91 & 480.65 & 3135.56 & 20564m & 10057k & 2847k \\
{\scshape Thai} & 1 & 30863.80 & 887.71 & 31751.51 & 61536m & 80499k & 8433k \\
        \toprule
    \end{tabular}
    }
 \caption{{\scshape Emu} 1481×2609×1843 and {\scshape Thai} $10240^3$ models: just each $k$th grid cell was used to reduce medial points count, meshing used the entire TSDF.}
 \label{tab:TSDF}
\end{table}

\begin{figure}[h]
    \centering
    \begin{adjustbox}{max width=\linewidth}
        \includegraphics{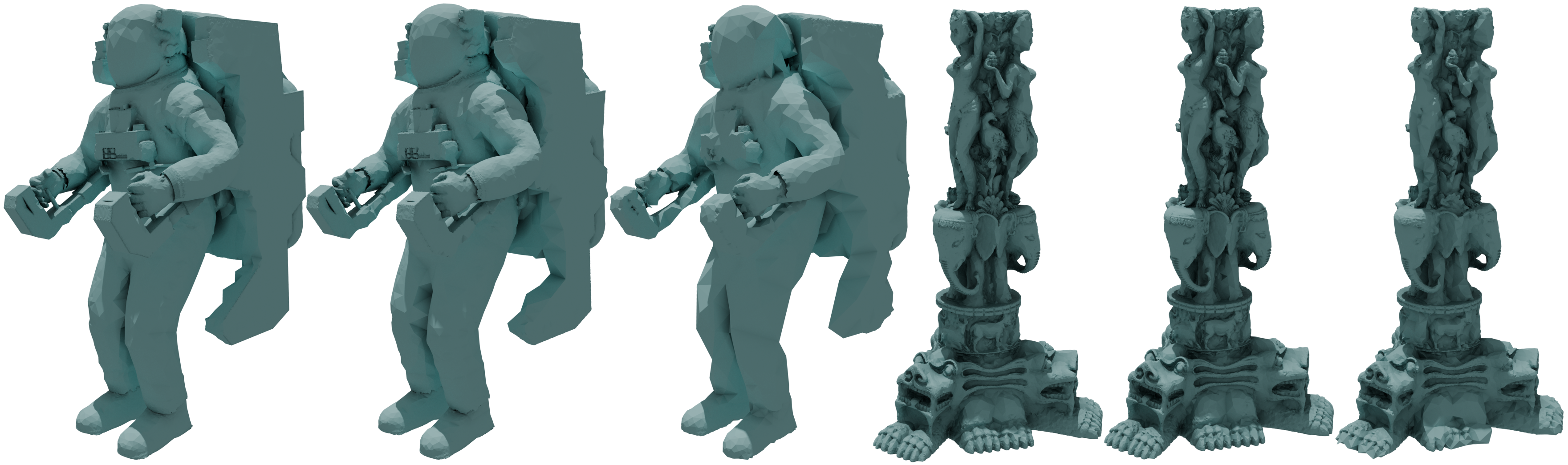}
    \end{adjustbox}
    \Description{A visualization showing results for different k values.}
    \caption{Results of FOMOS on huge input - meshes reconstructed with $\varepsilon=0.5$ from TSDFs \cite{openvdb} of size $1481 \times 2609 \times 1843$ {\scshape EMU} and $10240^3$ {\scshape Thai Statue}, for increasing $k=1,4,16$.}
    \label{fig:TSDF}
\end{figure}

Our algorithm does not rely on a regular grid but can place footpoints arbitrarily on $H$, as it does for SDFs defined on a grid near the surface.
As such, it is already well-suited for Truncated SDFs (TSDFs), sparse voxel grids containing SDF values only near the zero surface, thereby drastically reducing memory requirements.
To evaluate $\phi$ at a point $p$ far from the surface, we construct a kd-tree of the TSDF's grid corner values based on their locations, query $nn(p)$, and then add $\phi(nn(p))$ to $|p,nn(p)|$ as an approximation.
As a side benefit, our asymptotic time and space complexity becomes quadratic, since all operations are on surface rather than volume voxels (for non-sparse SDFs on a grid, the step of filtering the near-surface values is very fast but still cubic).
To quickly extract mesh previews from these huge inputs (as shown in Figure \ref{fig:TSDF}), we added a filter computing a subset of medial points (one per $k^{th}$ grid cell) while still meshing using the entire TSDF, cutting both memory and runtime from hours to seconds (see Table \ref{tab:TSDF}) without needing any large computing resources (AMD Ryzen 7 3.5GHz 16-core CPU).

\section{Experiments}

\label{experimentssec}

\begin{figure*}
    \centering
    \includegraphics[width=\linewidth]{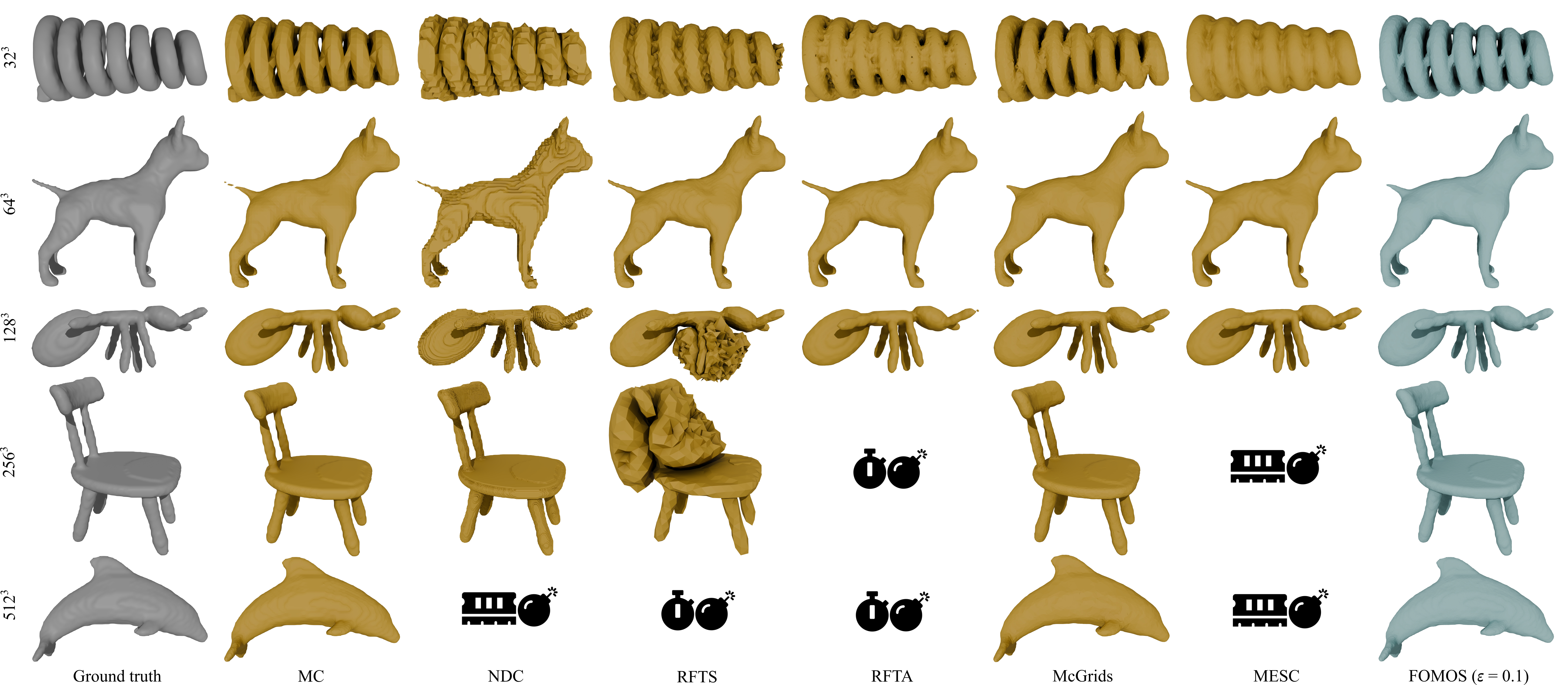}
    \caption{\textcolor{black}{Sample meshes from our benchmark reconstructed using various state-of-the-art methods and Ours, on varying grid sizes.}}
    \label{fig:compmodels}
\end{figure*}

Experimental evaluations were performed on an NVIDIA A100 GPU workstation (64 CPU cores, 64 GB RAM). This setup was chosen to accommodate the GPU requirements of the NDC, thereby providing a consistent computational baseline for all comparative analyses, despite our method’s independence from GPU hardware. We implemented our code in parallel on the CPU and used the 3D mesher component of CGAL \cite{cgal:rty-m3-24a,rineau2007generic}.

Our experimental dataset consists of 100 SDFs, sampled on grids from randomly selected shapes within the 
McGill dataset \cite{siddiqi2008retrieving}, including high-genus and thin-shell structures. Our method was extensively compared with various state-of-the-art SDF meshing techniques on our dataset: Marching Cubes \cite{chernyaev1995marching}, Neural Dual Contouring \cite{chen2022neural}, Reach for the Spheres \cite{sellan2023reach}, Reach for the Arcs \cite{sellan2024reach}, and Maximal Empty Sphere Contouring \cite{kohlbrenner2025polyhedral}. To prioritize efficient meshing, we imposed a 30-minute execution cap, after which the process was terminated. Figure \ref{fig:compmodels} illustrates results across varying grid resolutions. Notably, while Marching Cubes, McGrids, and FOMOS scaled effectively, other methods failed on larger grid sizes due to memory or time constraints.

\begin{table}[htb]
\centering
\resizebox{\linewidth}{!}{%
\begin{tabular}{lrrrrr}
    \toprule
    & \multicolumn{5}{c}{\textbf{Grid Size}} \\
    \cmidrule(lr){2-6}
    \textbf{Method}
    & \textbf{$32^3$}
    & \textbf{$64^3$}
    & \textbf{$128^3$}
    & \textbf{$256^3$}
    & \textbf{$512^3$} \\
    \midrule
    \rowcolor{lightgray} MC
    & \underline{0.000329}
    & \textbf{0.000077}
    & 0.000030
    & \textbf{0.000019}
    & \textbf{0.000017} \\

    NDC
    & 0.000410
    & 0.000092
    & 0.000034
    & \underline{0.000020}
    & $\times$ \\

    \rowcolor{lightgray} RFTS
    & 0.042924
    & 0.010607
    & 0.010054
    & 0.005694
    & $\times$ \\

    RFTA
    & 0.005068
    & 0.005587
    & \underline{0.000040}
    & 0.003288
    & $\times$ \\

    \rowcolor{lightgray} McGrids
    & 0.002787
    & 0.000098
    & 0.000028
    & 0.000028
    & 0.000026 \\

    MESC
    & \textbf{0.000091}
    & 0.002331
    & 0.001456
    & $\times$
    & $\times$ \\

    \rowcolor{lightgray} FOMOS ($\varepsilon=0.1$)
    & 0.001441
    & \underline{0.000090}
    & \textbf{0.000027}
    & 0.000026
    & \underline{0.000023} \\

    FOMOS ($\varepsilon=0.2$)
    & 0.004590
    & 0.000108
    & 0.000037
    & 0.000033
    & 0.000028 \\

    \rowcolor{lightgray} FOMOS ($\varepsilon=0.4$)
    & 0.010600
    & 0.000188
    & 0.000087
    & 0.000058
    & 0.000040 \\

    FOMOS ($\varepsilon=0.8$)
    & 0.017400
    & 0.000441
    & 0.000244
    & 0.000152
    & 0.000081 \\
    \bottomrule
\end{tabular}
}
\caption{Chamfer distance comparison across different SDF grid resolutions. Best results are shown in bold and second-best results are underlined. Missing values are marked with $\times$.}
\label{tab:chamfer_results}
\end{table}

\begin{table}[htb]
\centering
\resizebox{\linewidth}{!}{%
\begin{tabular}{lrrrrr}
    \toprule
    & \multicolumn{5}{c}{\textbf{Grid Size}} \\
    \cmidrule(lr){2-6}
    \textbf{Method}
    & \textbf{$32^3$}
    & \textbf{$64^3$}
    & \textbf{$128^3$}
    & \textbf{$256^3$}
    & \textbf{$512^3$} \\
    \midrule
    \rowcolor{lightgray} MC
    & 0.01053
    & 0.00556
    & 0.00354
    & $\underline{0.00261}$
    & \textbf{0.00217} \\

    NDC
    & 0.01225
    & 0.00623
    & 0.00378
    & 0.00267
    & $\times$ \\

    \rowcolor{lightgray} RFTS
    & 0.10170
    & 0.03043
    & 0.03047
    & 0.01917
    & $\times$ \\

    RFTA
    & 0.00909
    & 0.00767
    & $\underline{0.00309}$
    & 0.02115
    & $\times$ \\

    \rowcolor{lightgray} McGrids
    & 0.00617
    & $\mathbf{0.00364}$
    & 0.00322
    & 0.00315
    & 0.00310 \\

    MESC
    & \textbf{0.00485}
    & 0.00781
    & 0.00511
    & $\times$
    & $\times$ \\

    \rowcolor{lightgray} FOMOS ($\varepsilon=0.1$)
    & 0.01098
    & $\underline{0.00368}$
    & \textbf{0.00280}
    & \textbf{0.00256}
    & $\underline{0.00226}$ \\

    FOMOS ($\varepsilon=0.2$)
    & 0.02011
    & 0.00455
    & 0.00353
    & 0.00316
    & 0.00269 \\

    \rowcolor{lightgray} FOMOS ($\varepsilon=0.4$)
    & 0.03435
    & 0.00656
    & 0.00514
    & 0.00439
    & 0.00352 \\

    FOMOS ($\varepsilon=0.8$)
    & 0.05100
    & 0.01037
    & 0.00820
    & 0.00683
    & 0.00515 \\
    \bottomrule
\end{tabular}
}
\caption{Mean distance comparison across different SDF grid resolutions. Best results are shown in bold and second-best results are underlined. Missing values are marked with $\times$.}
\label{tab:mean_distance_results}
\end{table}

\begin{table}[htb]
\centering
\resizebox{\linewidth}{!}{%
\begin{tabular}{lrrrrr}
    \toprule
    & \multicolumn{5}{c}{\textbf{Grid Size}} \\
    \cmidrule(lr){2-6}
    \textbf{Method}
    & \textbf{$32^3$}
    & \textbf{$64^3$}
    & \textbf{$128^3$}
    & \textbf{$256^3$}
    & \textbf{$512^3$} \\
    \midrule
    \rowcolor{lightgray} MC
    & 0.00992
    & 0.00551
    & 0.00355
    & \underline{0.00261}
    & \underline{0.00217} \\

    NDC
    & 0.01128
    & 0.00606
    & 0.00378
    & 0.00266
    & $\times$ \\

    \rowcolor{lightgray} RFTS
    & 0.10038
    & 0.02843
    & 0.02808
    & 0.01781
    & $\times$ \\

    RFTA
    & 0.00885
    & 0.00774
    & \underline{0.00293}
    & 0.01876
    & $\times$ \\

    \rowcolor{lightgray} McGrids
    & 0.00617
    & \textbf{0.00327}
    & 0.00299
    & 0.00293
    & 0.00288 \\

    MESC
    & \textbf{0.00443}
    & 0.00750
    & 0.00510
    & $\times$
    & $\times$ \\

    \rowcolor{lightgray} FOMOS ($\varepsilon=0.1$)
    & \underline{0.00985}
    & \underline{0.00353}
    & \textbf{0.00268}
    & \textbf{0.00243}
    & \textbf{0.00216} \\

    FOMOS ($\varepsilon=0.2$)
    & 0.01936
    & 0.00429
    & 0.00332
    & 0.00298
    & 0.00256 \\

    \rowcolor{lightgray} FOMOS ($\varepsilon=0.4$)
    & 0.03053
    & 0.00603
    & 0.00477
    & 0.00411
    & 0.00333 \\

    FOMOS ($\varepsilon=0.8$)
    & 0.04563
    & 0.00949
    & 0.00753
    & 0.00637
    & 0.00485 \\
    \bottomrule
\end{tabular}
}
\caption{Median distance comparison across different SDF grid resolutions. Best results are shown in bold and second-best results are underlined. Missing values are marked with $\times$.}
\label{tab:median_distance_results}
\end{table}

\begin{table}[htb]
\centering
\resizebox{\linewidth}{!}{%
\begin{tabular}{lrrrrr}
    \toprule
    & \multicolumn{5}{c}{\textbf{Grid Size}} \\
    \cmidrule(lr){2-6}
    \textbf{Method}
    & \textbf{$32^3$}
    & \textbf{$64^3$}
    & \textbf{$128^3$}
    & \textbf{$256^3$}
    & \textbf{$512^3$} \\
    \midrule
    \rowcolor{lightgray} MC
    & 0.05322
    & \underline{0.02116}
    & \underline{0.01274}
    & \textbf{0.01140}
    & \textbf{0.01156} \\

    NDC
    & 0.04993
    & 0.02081
    & 0.01276
    & \underline{0.01144}
    & $\times$ \\

    \rowcolor{lightgray} RSTS
    & 0.39638
    & 0.14064
    & 0.14272
    & 0.09344
    & $\times$ \\

    RFTA
    & \underline{0.04208}
    & 0.03609
    & 0.01794
    & 0.09970
    & $\times$ \\

    \rowcolor{lightgray} McGrids
    & 0.11006
    & 0.02430
    & 0.01898
    & 0.01864
    & 0.01821 \\

    MESC
    & \textbf{0.02918}
    & 0.02596
    & 0.01602
    & $\times$
    & $\times$ \\

    \rowcolor{lightgray} FOMOS ($\varepsilon=0.1$)
    & 0.06384
    & \textbf{0.01480}
    & \textbf{0.01225}
    & 0.01189
    & \underline{0.01170} \\

    FOMOS ($\varepsilon=0.2$)
    & 0.08652
    & 0.01818
    & 0.01472
    & 0.01250
    & 0.01167 \\

    \rowcolor{lightgray} FOMOS ($\varepsilon=0.4$)
    & 0.14384
    & 0.03038
    & 0.02466
    & 0.01920
    & 0.01524 \\

    FOMOS ($\varepsilon=0.8$)
    & 0.20800
    & 0.05493
    & 0.04488
    & 0.03408
    & 0.02627 \\
    \bottomrule
\end{tabular}
}
\caption{Bidirectional Hausdorff distance comparison across different SDF grid resolutions. Best results are shown in bold and second-best results are underlined. Missing values are marked with $\times$.}
\label{tab:hausdorff_bidirectional_results}
\end{table}

\begin{table}[htb]
\centering
\resizebox{\linewidth}{!}{%
\begin{tabular}{lrrrrr}
    \toprule
    & \multicolumn{5}{c}{\textbf{Grid Size}} \\
    \cmidrule(lr){2-6}
    \textbf{Method}
    & \textbf{$32^3$}
    & \textbf{$64^3$}
    & \textbf{$128^3$}
    & \textbf{$256^3$}
    & \textbf{$512^3$} \\
    \midrule
    \rowcolor{lightgray} MC
    & 36.39001
    & 94.01181
    & \textbf{99.54573}
    & \underline{99.60668}
    & \underline{99.61534} \\

    NDC
    & 29.25581
    & 92.41970
    & 99.52385
    & \textbf{99.60988}
    & $\times$ \\

    \rowcolor{lightgray} RFTS
    & 10.66636
    & 62.64968
    & 59.87974
    & 70.66689
    & $\times$ \\

    RFTA
    & \textbf{95.61512}
    & 97.44285
    & 97.35788
    & 66.63557
    & $\times$ \\

    \rowcolor{lightgray} McGrids
    & 80.78618
    & 96.92392
    & 97.23969
    & 97.37428
    & 97.54298 \\

    MESC
    & \underline{91.62699}
    & 95.74503
    & 98.67265
    & $\times$
    & $\times$ \\

    \rowcolor{lightgray} FOMOS ($\varepsilon=0.1$)
    & 79.66533
    & \textbf{97.88816}
    & \underline{98.86375}
    & 98.80491
    & \textbf{98.97040} \\

    FOMOS ($\varepsilon=0.2$)
    & 65.06268
    & 96.35759
    & 98.45749
    & 98.74143
    & 98.82500 \\

    \rowcolor{lightgray} FOMOS ($\varepsilon=0.4$)
    & 35.04069
    & 83.67735
    & 90.73764
    & 95.90346
    & 98.58905 \\

    FOMOS ($\varepsilon=0.8$)
    & 9.58652
    & 48.93296
    & 61.33698
    & 72.74585
    & 90.69341 \\
    \bottomrule
\end{tabular}
}
\caption{F-Score comparison across different SDF grid resolutions. Best results are shown in bold and second-best results are underlined. Missing values are marked with $\times$.}
\label{tab:fscore_results}
\end{table}

\begin{table}[htb]
\centering
\resizebox{\linewidth}{!}{%
\begin{tabular}{lrrrrr}
    \toprule
    & \multicolumn{5}{c}{\textbf{Grid Size}} \\
    \cmidrule(lr){2-6}
    \textbf{Method}
    & \textbf{$32^3$}
    & \textbf{$64^3$}
    & \textbf{$128^3$}
    & \textbf{$256^3$}
    & \textbf{$512^3$} \\
    \midrule
    \rowcolor{lightgray} MC
    & 0.93159
    & 0.98360
    & \underline{0.99526}
    & \textbf{0.99795}
    & \textbf{0.99824} \\

    NDC
    & 0.89585
    & 0.95810
    & 0.97081
    & 0.97308
    & $\times$ \\

    \rowcolor{lightgray} RFTS
    & 0.63362
    & 0.88390
    & 0.87597
    & 0.93328
    & $\times$ \\

    RFTA
    & \textbf{0.97691}
    & \textbf{0.98916}
    & 0.99151
    & 0.87091
    & $\times$ \\

    \rowcolor{lightgray} McGrids
    & 0.90923
    & \underline{0.98597}
    & 0.99218
    & 0.99277
    & 0.99280 \\

    MESC
    & \underline{0.97261}
    & 0.97742
    & 0.99286
    & $\times$
    & $\times$ \\

    \rowcolor{lightgray} FOMOS ($\varepsilon=0.1$)
    & 0.92935
    & 0.98813
    & \textbf{0.99528}
    & \underline{0.99601}
    & 0.99481 \\

    FOMOS ($\varepsilon=0.2$)
    & 0.89749
    & 0.98783
    & \underline{0.99508}
    & 0.99680
    & \underline{0.99771} \\

    \rowcolor{lightgray} FOMOS ($\varepsilon=0.4$)
    & 0.82800
    & 0.97812
    & 0.98931
    & 0.99351
    & 0.99734 \\

    FOMOS ($\varepsilon=0.8$)
    & 0.74957
    & 0.95273
    & 0.97284
    & 0.98213
    & 0.99156 \\
    \bottomrule
\end{tabular}
}
\caption{Normal consistency comparison across different SDF grid resolutions. Best results are shown in bold and second-best results are underlined. Missing values are marked with $\times$.}
\label{tab:normal_consistency_results}
\end{table}


\paragraph{Evaluation on Meshing Quality} First, we evaluate several complementary metrics like global geometry, local outliers, robustness, and visual quality. Specifically, we compare results across: (1) distance metrics (chamfer, mean, median, RMSE, and Hausdorff) to quantify geometric accuracy in terms of typical, robust, and worst-case errors, ensuring both average fidelity and bounds on local deviations, (2) F-score emphasizing how much of the surface lies within a tolerance (0.005), and (3) Normal consistency, to evaluate preservation of local surface orientation (crucial for visual quality).
Results in Tables \ref{tab:chamfer_results} to \ref{tab:normal_consistency_results} confirm that FOMOS, except at the lowest $32^3$, 
remained better than or comparable to SOTA methods across all metrics plus topology preservation. 

\begin{figure}
    \centering
    \includegraphics[width=\linewidth]{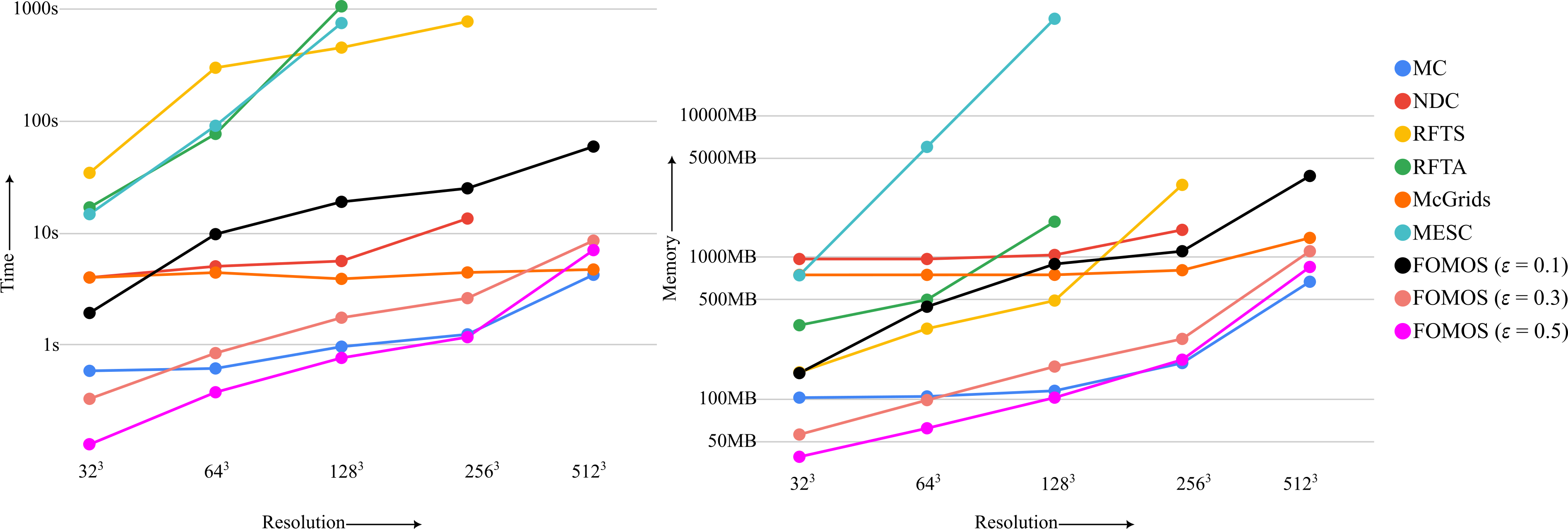}
    \caption{Runtime and memory comparison (both logarithmic scale).}
    \label{fig:timechart}
\end{figure}

\begin{figure*}
    \includegraphics[width=\linewidth]{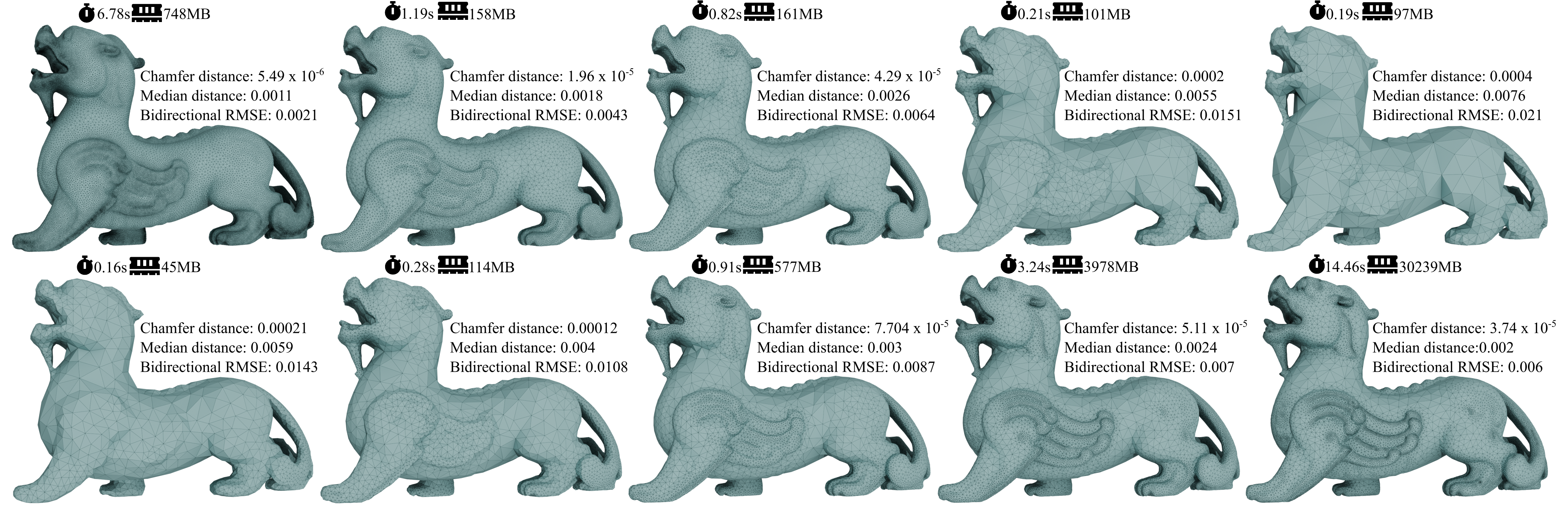}
    \caption{Effect of grid resolution and value of $\varepsilon$ on running time, memory and error metrics on {\scshape Mausoleum} model. Top row (from left to right): Grid Size $ =128^3$ and $\varepsilon$ as 0.1, 0.2, 0.3 (for 0.5, see $128^3$ in bottom row), 0.7, 1.0, Bottom row (from left to right): $\varepsilon =0.5$ and Grid size as $64^3, 128^3, 256^3, 512^3, 1024^3$ respectively. }
    \label{fig:resolution}
\end{figure*}

\paragraph{Runtime and memory requirement} The efficiency of the meshing algorithm is another important aspect, as it determines whether complex, high-resolution surfaces can be generated in real-time or if the process is bottlenecked by hardware limitations and slow processing cycles. This factor is particularly important for interactive setups, evolving physical simulation, or deployment on limited hardware such as a VR headset (for instance, \cite{Sureshkumar2025RibbonSculpt}). Figure \ref{fig:timechart} shows the execution time and memory consumption for various methods on a logarithmic scale. As shown, FOMOS, even with a computationally intensive parameter of $\varepsilon=0.1$, demonstrated superior efficiency compared to alternatives excluding MC and NDC (which failed to produce acceptable, high-quality surfaces). Furthermore, the memory requirements of FOMOS were also comparable to those of other techniques. Figure \ref{fig:resolution} shows how the memory and execution time vary with the $\varepsilon$ value, and that FOMOS maintains high efficiency in terms of both temporal and spatial complexity.
\textcolor{black}{Note that our algorithm's time and spatial complexity are fundamentally different from methods that only handle discrete SDFs. While their complexities are input-bound, a function of the (cubic) grid cell count, applying the mesher enables ours to become output-bound in the order of the (quadratic) target triangle count.}

\begin{figure}
    \centering
    \includegraphics[width=0.8\linewidth]{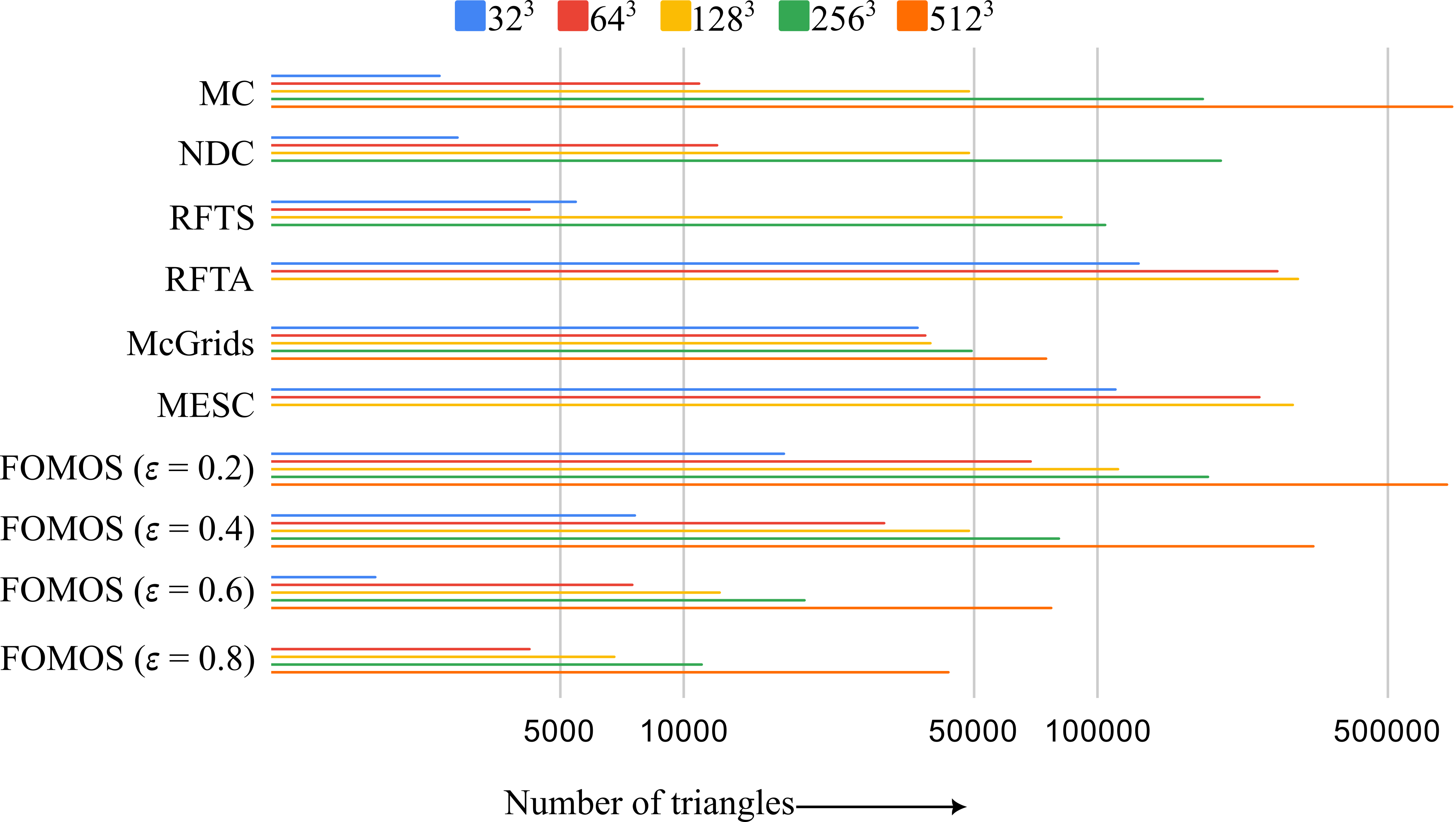}
    \caption{Number of triangles in meshes generated by different methods.}
    \label{fig:placeholder}
\end{figure}

\begin{figure}
    \centering
    \includegraphics[width=\linewidth]{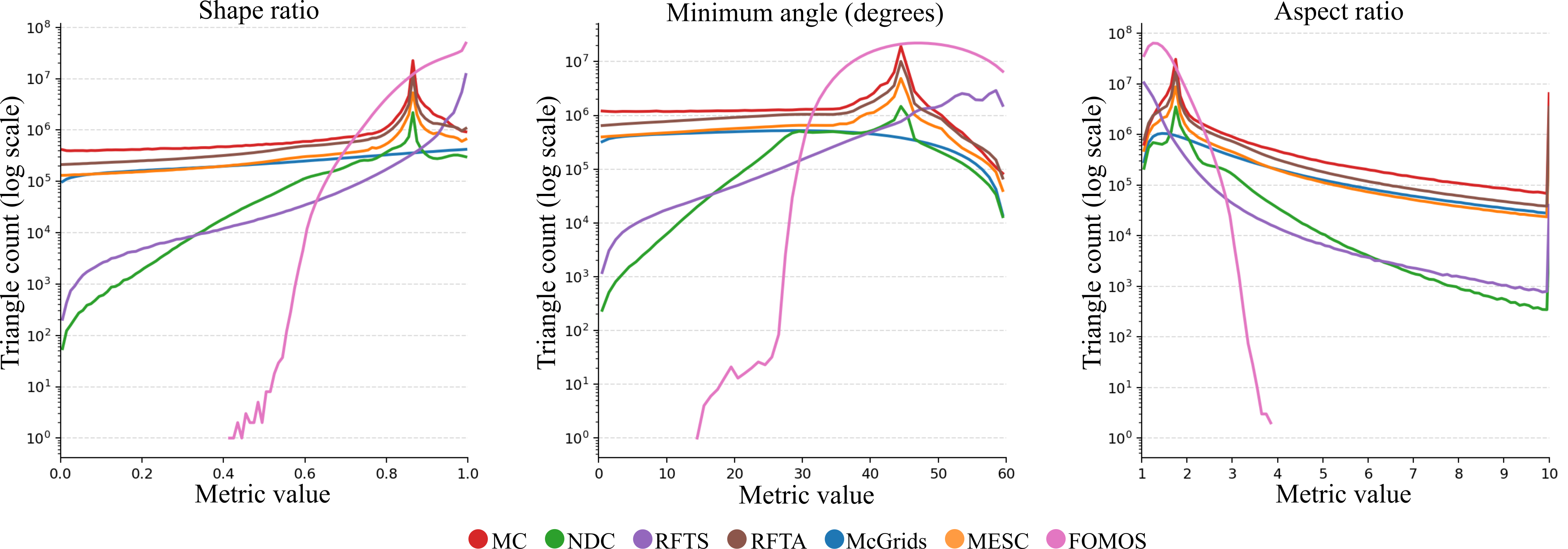}
    \caption{\textcolor{black}{Triangle-quality distributions. Curves show absolute triangle counts per bin on a logarithmic scale. For side lengths \(a,b,c\), area \(A\), longest side \(L=\max(a,b,c)\), and interior angles \(\theta_1,\theta_2,\theta_3\), we measure $Q_{\mathrm{shape}}=\frac{4\sqrt{3}\,A}{a^2+b^2+c^2}$, $Q_{\mathrm{aspect}}=\frac{\sqrt{3}\,L^2}{4A}$, $\theta_{\min}=\min_i\theta_i$
Equilateral triangles have \(Q_{\mathrm{shape}}=Q_{\mathrm{aspect}}=1\) and \(\theta_{\min}=60^\circ\). Higher shape quality and minimum angle, and lower aspect ratio, indicate better quality. Aspect ratios \(\geq10\) are grouped at 10. Counts are unnormalized and thus reflect both mesh size and triangle quality.}}
    \label{fig:triquality}
\end{figure}

\paragraph{Triangle quality} Triangle quality is critical for downstream applications, such as rendering and physical simulation. To make the reconstructed mesh readily usable, the triangles must be high-quality. We evaluate the resulting triangle quality using three metrics: (1) shape ratio, which compares how closely the triangle approximates an equilateral triangle; (2) minimum angle, an indicator of "sliver" triangles (thin, needle-like shapes); and (3) aspect ratio (the ratio of the longest edge to the shortest) that measures the "stretch" or elongation of the triangle. Figure \ref{fig:triquality} illustrates the comparison of these metrics on meshes generated by various methods, showing that the meshes created by FOMOS exhibit high-quality triangles in contrast to all other SOTA methods - high shape ratios (0.4-1.0), high minimum angles ($15^\circ -60 ^\circ$), and low aspect ratios (1-4).

\begin{table}[htb]
\centering
\resizebox{\linewidth}{!}{%
\begin{tabular}{lrrrr}
    \toprule
    \textbf{Method}
    & \textbf{Percent Defective} 
    & \textbf{Mean $\eta$}
    & \textbf{Min Angle ($^\circ$)}
    & \textbf{Max Angle ($^\circ$)} \\
    \midrule
    \rowcolor{lightgray} MC
    & 11.86301
    & 0.70920
    & 0.02256
    & 178.18326 \\

    NDC
    & 0.67634
    & 0.79736
    & 2.22531
    & 174.12696 \\

    \rowcolor{lightgray} RFTS
    & \underline{1.34738}
    & \textbf{0.92457}
    & \underline{13.01416}
    & \underline{146.92335} \\

    RFTA
    & 10.21209
    & 0.71708
    & 0.00134
    & 170.34571 \\

    \rowcolor{lightgray} McGrids
    & 17.85112
    & 0.61417
    & 0.00601
    & 179.38640 \\

    MES
    & 10.38297
    & 0.71411
    & 0.00133
    & 172.50748 \\

    \rowcolor{lightgray} FOMOS
    & \textbf{0.00000}
    & \underline{0.92297}
    & \textbf{27.11516}
    & \textbf{122.57808} \\
    \bottomrule
\end{tabular}%
}
\caption{\textcolor{black}{Triangle quality comparison across reconstruction methods. Lower Percent Defective and Max Angle values, and higher Mean $\eta$ and Min Angle values, indicate better triangle quality. Best results are shown in bold and second-best results are underlined.}}
\label{tab:triangle_quality}
\end{table}

\paragraph{Dependence on grid resolution and $\varepsilon$ parameter} Our primary comparisons used $\varepsilon=0.1$ for very fine resolutions, but that parameter can range from $0<\varepsilon<1$. While FOMOS provides a more accurate approximation for lower $\varepsilon$ values, this comes at the cost of increased memory usage, longer execution time, and more triangles. At the same time, larger $\varepsilon$ values allow sparse approximations with less memory and time. Similarly, as expected, increasing the grid resolution improves the reconstruction accuracy. Tables \ref{tab:chamfer_results} to \ref{tab:normal_consistency_results} 
compares error profiles across various $\varepsilon$ values, and Figure \ref{fig:resolution} compares the scaling of time and memory requirements with respect to the grid resolution and $\varepsilon$ value. As demonstrated, the lower the value of epsilon, the higher the reconstruction accuracy is. Note that the share of the medial point computation runtime increases with higher resolution, while the share of meshing increases with smaller epsilon values.
\textcolor{black}{A suitable grid resolution can be chosen by the user based on the properties of the underlying continuous SDF, namely the finest features that they desire to recover, while FOMOS adapts its sampling density automatically proportionally to features, with $\varepsilon$ determining how dense the features themselves are sampled. This automated determination of adaptive sampling density also works for noisy input if the noise is small enough compared to the features.}

\begin{figure}
    \centering
    \includegraphics[width=\linewidth]{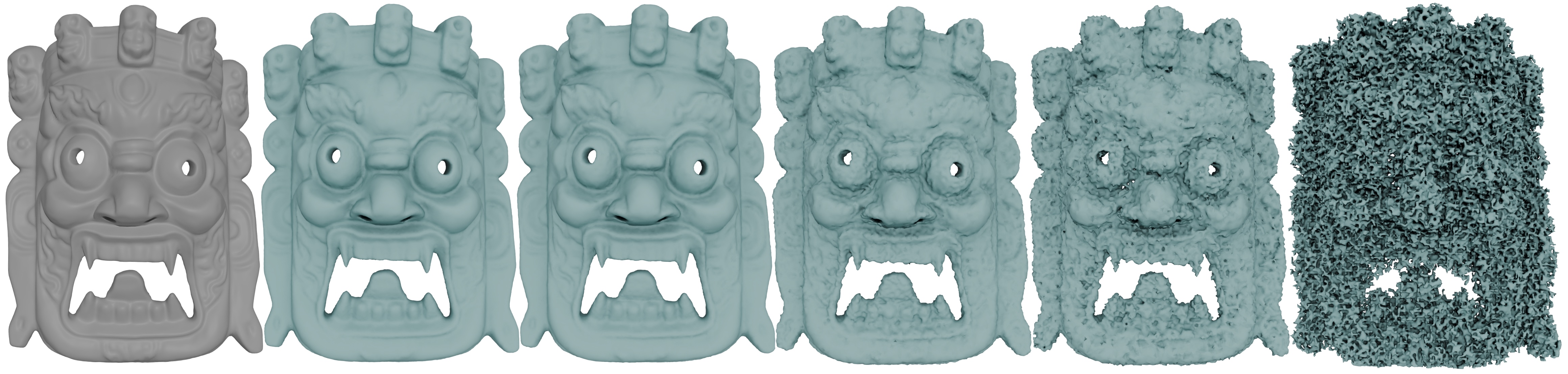}
    \caption{Effect of noise on reconstruction on a $128^3$ grid, with $\varepsilon=0.3$. Left to right: Ground-truth mesh, results of FOMOS for $\eta$=0, 0.001, 0.005, 0.01, 0.05 noise extent of the bounding box diagonal.}
    \label{fig:noise}
\end{figure}

\begin{table*}[htb]
\centering
\resizebox{0.8\linewidth}{!}{%
\begin{tabular}{
    l
    rrrrr
    @{\hspace{6pt}\vrule\hspace{6pt}}rrrrr
    @{\hspace{6pt}\vrule\hspace{6pt}}rrrrr
}
    \toprule
    & \multicolumn{5}{c}{\textbf{Global Topology Accuracy}}
    & \multicolumn{5}{c}{\textbf{Perfect Match Accuracy (B0)}}
    & \multicolumn{5}{c}{\textbf{Perfect Match Accuracy (B1)}} \\
    \cmidrule(lr){2-6}
    \cmidrule(lr){7-11}
    \cmidrule(lr){12-16}

    \textbf{Method}
    & \textbf{$32^3$} & \textbf{$64^3$} & \textbf{$128^3$}
    & \textbf{$256^3$} & \textbf{$512^3$}
    & \textbf{$32^3$} & \textbf{$64^3$} & \textbf{$128^3$}
    & \textbf{$256^3$} & \textbf{$512^3$}
    & \textbf{$32^3$} & \textbf{$64^3$} & \textbf{$128^3$}
    & \textbf{$256^3$} & \textbf{$512^3$} \\
    \midrule

    \rowcolor{lightgray}
    MC
    & 68 & 84 & 81 & 68 & 60
    & 82 & 96 & 90 & 87 & 72
    & 83 & 87 & 84 & 71 & 65 \\

    NDC
    & 67 & 80 & 78 & 70 & $\times$
    & 92 & 99 & 97 & 95 & $\times$
    & 72 & 81 & 79 & 71 & $\times$ \\

    \rowcolor{lightgray}
    RTFS
    & \underline{84} & 86 & 85 & 87.88 & $\times$
    & \textbf{100} & \textbf{100} & \textbf{100} & \textbf{100} & $\times$
    & 84 & 86 & 85 & 87.88 & $\times$ \\

    RTFA
    & 6 & 79 & 57 & $\times$ & $\times$
    & 9 & 80 & 57 & $\times$ & $\times$
    & 79 & 95 & 96 & $\times$ & $\times$ \\

    \rowcolor{lightgray}
    McGrids
    & 63.92 & 88.14 & 94.85 & 96.92 & 94.36
    & 98.45 & \textbf{100} & \textbf{100} & \textbf{100} & \textbf{100}
    & 64.43 & 88.14 & 94.85 & 96.92 & 94.36 \\

    MESC
    & 83 & 89 & 96 & $\times$ & $\times$
    & 97 & 93 & 97 & $\times$ & $\times$
    & 84 & 95 & 97 & $\times$ & $\times$ \\

    \rowcolor{lightgray}
    FOMOS
    & 83.84 & \textbf{97.98} & \textbf{97.98} & \textbf{97.98} & \textbf{100}
    & 98.99 & \textbf{100} & \textbf{100} & \textbf{100} & \textbf{100}
    & \textbf{84.85} & \textbf{97.98} & \textbf{97.98} & \textbf{97.98} & \textbf{100} \\
    \bottomrule
\end{tabular}
}
\caption{Global topology and per-dimension Betti-number perfect match accuracy across SDF grid resolutions. Best results are shown in bold and second-best results are underlined. Missing values are marked with $\times$.}
\label{tab:topology_accuracy}
\end{table*}

\color{black}
\paragraph{Topological correctness} We evaluate the topological fidelity of reconstructed meshes using global topology accuracy and perfect-match accuracy for the zeroth and first Betti numbers, \(B_0\) and \(B_1\), which characterize the number of connected components and independent loops, respectively. Global topology accuracy measures the proportion of reconstructions that exactly match the reference topology, while dimension-wise metrics assess agreement for each invariant. As shown in Table~\ref{tab:topology_accuracy}, FOMOS achieves the highest global topology accuracy among the evaluated methods at every tested resolution, increasing from \(83.84\%\) at \(32^3\) to \(97.98\%\) at \(64^3\), and maintaining this accuracy at \(128^3\) and \(256^3\) before reaching \(100\%\) at \(512^3\). Its dimension-wise results are similarly strong: \(B_0\) accuracy reaches \(100\%\) from \(64^3\) onward, while \(B_1\) accuracy reaches \(97.98\%\) over the same range of intermediate resolutions and \(100\%\) at \(512^3\). In comparison, McGrids improves substantially with increasing resolution, reaching \(96.92\%\) global accuracy at \(256^3\), but remains below FOMOS at the evaluated resolutions. Other methods also perform well on individual invariants; for example, RoS achieves \(100\%\) \(B_0\) accuracy up to \(256^3\), yet its global accuracy remains lower. These results demonstrate that matching individual Betti numbers does not necessarily guarantee recovery of the complete reference topology. Overall, FOMOS provides the strongest global topological agreement in our experiments, with its accuracy improving consistently across the evaluated resolutions and reaching perfect agreement at \(512^3\).
\color{black}

\paragraph{Effect of noise} We also experimented with the reconstruction after injecting noise by perturbing $\phi$ at each grid location within a range $\eta$ (noise extent) of the bounding box diagonal.
The impact of this noise on our reconstruction quality can be seen in Figure \ref{fig:noise}. It is worth noting that FOMOS could capture the overall shape even for a very high level of noise.

\begin{table*}[htb]
\centering
\resizebox{\linewidth}{!}{%
\begin{tabular}{ lcrrrr rrrr rrrr rrrr }
    \toprule
    \multirow{2}{*}{\textbf{Dataset}} & \multirow{2}{*}{\textbf{Res.}} 
      & \multicolumn{4}{c}{\textbf{Chamfer Distance} ($\downarrow$)} 
      & \multicolumn{4}{c}{\textbf{RMSE} ($\downarrow$)}
      & \multicolumn{4}{c}{\textbf{F-Score (\%)} ($\uparrow$)} 
      & \multicolumn{4}{c}{\textbf{Normal Consistency} ($\uparrow$)} \\
    \cmidrule(lr){3-6} \cmidrule(lr){7-10} \cmidrule(lr){11-14} \cmidrule(lr){15-18}
     & \textbf{$\varepsilon$} & \textbf{0.1} & \textbf{0.2} & \textbf{0.4} & \textbf{0.8} 
       & \textbf{0.1} & \textbf{0.2} & \textbf{0.4} & \textbf{0.8} 
       & \textbf{0.1} & \textbf{0.2} & \textbf{0.4} & \textbf{0.8} 
       & \textbf{0.1} & \textbf{0.2} & \textbf{0.4} & \textbf{0.8} \\
    \midrule
    \rowcolor{lightgray} \cellcolor{white}
    & 512 & $1.61\times 10^{-5}$ & $1.8\times 10^{-5}$ & $2.8\times 10^{-5}$ & $6.8\times 10^{-5}$ & 0.0037 & 0.0036 & 0.0038 & 0.0065 & 99.51 & 99.54 & 99.38 & 92.04 & 0.9947 & 0.9962 & 0.9959 & 0.9899 \\
    & 256 & $1.7\times 10^{-5}$ & $2.3\times 10^{-5}$ & $4.6\times 10^{-5}$ & $1.36\times 10^{-4}$ & 0.0036 & 0.0036 & 0.0049 & 0.0101 & 99.68 & 99.65 & 97.15 & 73.74 & 0.9955 & 0.9961 & 0.9931 & 0.9811 \\
    \rowcolor{lightgray} \cellcolor{white}
    & 128 & $1.9\times 10^{-5}$ & $2.9\times 10^{-5}$ & $6.8\times 10^{-5}$ & $2.18\times 10^{-4}$ & 0.0036 & 0.0038 & 0.0063 & 0.0131 & 99.69 & 99.33 & 91.94 & 61.76 & 0.9954 & 0.9952 & 0.9902 & 0.9728 \\
    & 64  & $2.3\times 10^{-5}$ & $4.0\times 10^{-5}$ & $1.0\times 10^{-4}$ & $3.39\times 10^{-4}$ & 0.0038 & 0.0045 & 0.008 & 0.0165 & 99.58 & 98.1 & 86.21 & 49.95 & 0.9919 & 0.9919 & 0.985 & 0.9604 \\
    \rowcolor{lightgray} \cellcolor{white}
    \multirow{-5}{*}{\textbf{McGill}} 
    & 32  & $6.4\times 10^{-4}$ & $2.8\times 10^{-3}$ & $6.9\times 10^{-3}$ & $1.27\times 10^{-2}$ & 0.0098 & 0.0221 & 0.046 & 0.0767 & 92.37 & 82.92 & 52.58 & 17.45 & 0.9594 & 0.935 & 0.8803 & 0.8004 \\
    \midrule
    \rowcolor{lightgray} \cellcolor{white}
    & 512 & $6.1\times 10^{-5}$ & $7.2\times 10^{-5}$ & $8.0\times 10^{-5}$ & $1.0\times 10^{-4}$ & 0.0066 & 0.0067 & 0.0064 & 0.0064 & 95.14 & 94.66 & 94.46 & 93.3 & 0.9863 & 0.9932 & 0.9932 & 0.9854 \\
    & 256 & $7.1\times 10^{-5}$ & $7.7\times 10^{-5}$ & $9.0\times 10^{-5}$ & $1.4\times 10^{-4}$ & 0.0064 & 0.0063 & 0.0062 & 0.0082 & 95.64 & 95.46 & 94.93 & 85.3 & 0.989 & 0.992 & 0.9871 & 0.9714 \\
    \rowcolor{lightgray} \cellcolor{white}
    & 128 & $6.1\times 10^{-5}$ & $6.7\times 10^{-5}$ & $8.0\times 10^{-5}$ & $1.28\times 10^{-4}$ & 0.0062 & 0.006 & 0.0059 & 0.0080 & 95.95 & 95.76 & 95.23 & 85.55 & 0.989 & 0.992 & 0.987 & 0.971 \\
    & 64  & $3.4\times 10^{-4}$ & $9.8\times 10^{-4}$ & $2.2\times 10^{-3}$ & $3.8\times 10^{-3}$ & 0.0079 & 0.0115 & 0.0209 & 0.0346 & 92.7 & 90.51 & 79.62 & 46.38 & 0.9526 & 0.9468 & 0.9264 & 0.8841 \\
    \rowcolor{lightgray} \cellcolor{white}
    \multirow{-5}{*}{\textbf{SHREC}} 
    & 32  & $6.69\times 10^{-3}$ & $8.6\times 10^{-3}$ & $1.3\times 10^{-2}$ & $1.9\times 10^{-2}$ & 0.0336 & 0.0506 & 0.0787 & 0.107 & 73.34 & 64.2 & 38.01 & 11.67 & 0.8558 & 0.8355 & 0.7822 & 0.7136 \\
    \midrule
    \rowcolor{lightgray} \cellcolor{white}
    & 512 & $6.5\times 10^{-3}$ & $6.9\times 10^{-3}$ & $7.3\times 10^{-3}$ & $8.5\times 10^{-3}$ & 0.0661 & 0.0679 & 0.0692 & 0.071 & 40.72 & 37.09 & 34.78 & 32.08 & 0.9613 & 0.9722 & 0.9705 & 0.9572 \\
    & 256 & $7.6\times 10^{-3}$ & $7.9\times 10^{-3}$ & $8.5\times 10^{-3}$ & $1.1\times 10^{-2}$ & 0.071 & 0.073 & 0.073 & 0.075 & 29.15 & 27.26 & 26.18 & 23.81 & 0.9688 & 0.9741 & 0.965 & 0.9443 \\
    \rowcolor{lightgray} \cellcolor{white}
    & 128 & $8.4\times 10^{-3}$ & $8.9\times 10^{-3}$ & $1.04\times 10^{-2}$ & $1.4\times 10^{-2}$ & 0.075 & 0.0755 & 0.0765 & 0.0796 & 22.29 & 21.62 & 20.63 & 17.2 & 0.967 & 0.9623 & 0.9466 & 0.9105 \\
    & 64  & $9.5\times 10^{-3}$ & $1.2\times 10^{-2}$ & $2.7\times 10^{-2}$ & $4.4\times 10^{-2}$ & 0.0789 & 0.0795 & 0.083 & 0.0909 & 19.83 & 19.22 & 17.09 & 11.55 & 0.9458 & 0.9378 & 0.9096 & 0.8547 \\
    \rowcolor{lightgray} \cellcolor{white}
    \multirow{-5}{*}{\textbf{Thingi10K}} 
    & 32  & $9.74\times 10^{-3}$ & $1.32\times 10^{-2}$ & $2.67\times 10^{-2}$ & $4.36\times 10^{-2}$ & 0.0732 & 0.0819 & 0.1105 & 0.1448 & 19.11 & 16.32 & 10.94 &  4.73 & 0.8652 & 0.8410 & 0.7877 & 0.7342 \\
    \bottomrule
\end{tabular}%
}
\caption{\textcolor{black}{Reconstruction metrics across datasets, resolutions, and $\varepsilon$, reporting Chamfer distance, RMSE, F-Score, and Normal Consistency.}}
\label{tab:unified_reconstruction_results}

\end{table*}

{\paragraph{Evaluation on different datasets} A comprehensive evaluation was conducted on the McGill, SHREC, and a subset of 1,000 randomly selected meshes from the Thingi10K dataset to assess the robustness of FOMOS across geometries of increasing complexity, resulting in a total of 40,000 reconstructions. These datasets were deliberately selected because they represent distinct reconstruction challenges. McGill contains smooth organic shapes, making it suitable for evaluating reconstruction accuracy under favorable conditions. SHREC contains thin structures and complex topologies that are particularly sensitive to discretization and sampling errors. In contrast, Thingi10K contains CAD-like objects with sharp edges, corners, and planar regions that are inherently challenging for SDF meshing algorithms. Table \ref{tab:unified_reconstruction_results} summarizes the quantitative evaluation. Across all datasets, the proposed method consistently achieves its best performance at lower $\varepsilon$ values and higher grid resolutions, as evidenced by lower Chamfer distances and RMSE values together with higher F-score and normal consistency. Although minor non-monotonic variations are observed for certain metric-parameter combinations, these fluctuations are expected due to the competing effects of detail preservation, numerical noise, and isosurface extraction artifacts, as well as the differing sensitivities of the evaluation metrics. The results further reveal that the performance gap between McGill and SHREC is relatively modest, suggesting that the proposed method can accurately reconstruct thin structures and complex topologies. The substantially larger errors observed on Thingi10K are primarily attributable to the prevalence of sharp CAD features, which cannot be perfectly represented by a smooth implicit reconstruction and therefore tend to be rounded during surface extraction. Nevertheless, the consistent improvement in all metrics as $\varepsilon$ decreases and resolution increases demonstrates that the proposed algorithm effectively captures geometric detail and converges toward the target surface, while maintaining strong performance across a diverse range of shape classes and reconstruction challenges.}

\begin{table}[t]
\centering
\resizebox{0.8\linewidth}{!}{%
\begin{tabular}{cccccc}
\toprule
Resolution & $\varepsilon$ &
\multicolumn{2}{c}{Chamfer Distance (CD) $\downarrow$} &
\multicolumn{2}{c}{Hausdorff Distance (HD) $\downarrow$} \\
\cmidrule(lr){3-4}\cmidrule(lr){5-6}
& & FOMOS & Curvature & FOMOS & Curvature \\
\midrule

\multirow{4}{*}{64}
& \cellcolor{lightgray}0.1
& \cellcolor{lightgray}${9.95\times10^{-5}}$
& \cellcolor{lightgray}${2.1\times10^{-4}}$
& \cellcolor{lightgray}${0.0363}$
& \cellcolor{lightgray}${0.0629}$ \\
& 0.2
& $1.75\times10^{-4}$
& $3.15\times10^{-4}$
& ${0.0499}$
& ${0.0724}$ \\
& \cellcolor{lightgray}0.4
& \cellcolor{lightgray}${5.07\times10^{-4}}$
& \cellcolor{lightgray}$6.82\times10^{-4}$
& \cellcolor{lightgray}${0.0779}$
& \cellcolor{lightgray}${0.1057}$ \\
& 0.8
& ${1.65\times10^{-3}}$
& $1.62\times10^{-3}$
& ${0.1292}$
& ${0.1278}$ \\
\midrule

\multirow{4}{*}{128}
& \cellcolor{lightgray}0.1
& \cellcolor{lightgray}$6.42\times10^{-5}$
& \cellcolor{lightgray}${1.56\times10^{-4}}$
& \cellcolor{lightgray}${0.023}$
& \cellcolor{lightgray}${0.0498}$ \\
& 0.2
& ${9.98\times10^{-5}}$
& $2.46\times10^{-4}$
& ${0.0292}$
& ${0.064}$ \\
& \cellcolor{lightgray}0.4
& \cellcolor{lightgray}${2.29\times10^{-4}}$
& \cellcolor{lightgray}$5.39\times10^{-4}$
& \cellcolor{lightgray}${0.0469}$
& \cellcolor{lightgray}${0.0909}$ \\
& 0.8
& ${7.3\times10^{-4}}$
& $1.43\times10^{-3}$
& ${0.083}$
& ${0.1244}$ \\
\midrule

\multirow{4}{*}{256}
& \cellcolor{lightgray}0.1
& \cellcolor{lightgray}${6.56\times10^{-5}}$
& \cellcolor{lightgray}$1.47\times10^{-4}$
& \cellcolor{lightgray}${0.0224}$
& \cellcolor{lightgray}${0.0458}$ \\
& 0.2
& ${8.63\times10^{-5}}$
& $2.12\times10^{-4}$
& ${0.0236}$
& ${0.0584}$ \\
& \cellcolor{lightgray}0.4
& \cellcolor{lightgray}${1.65\times10^{-4}}$
& \cellcolor{lightgray}$4.79\times10^{-4}$
& \cellcolor{lightgray}${0.0348}$
& \cellcolor{lightgray}${0.0863}$ \\
& 0.8
& ${4.79\times10^{-4}}$
& $1.1\times10^{-3}$
& ${0.0621}$
& ${0.1167}$ \\
\midrule

\multirow{4}{*}{512}
& \cellcolor{lightgray}0.1
& \cellcolor{lightgray}${6.55\times10^{-5}}$
& \cellcolor{lightgray}$1.32\times10^{-4}$
& \cellcolor{lightgray}${0.0206}$
& \cellcolor{lightgray}$0.0448$ \\
& 0.2
& ${7.75\times10^{-5}}$
& $1.72\times10^{-4}$
& ${0.02}$
& $0.0552$ \\
& \cellcolor{lightgray}0.4
& \cellcolor{lightgray}${1.16\times10^{-4}}$
& \cellcolor{lightgray}$3.42\times10^{-4}$
& \cellcolor{lightgray}${0.0269}$
& \cellcolor{lightgray}$0.0746$ \\
& 0.8
& ${2.63\times10^{-4}}$
& $7.48\times10^{-4}$
& ${0.0474}$
& $0.1003$ \\
\bottomrule
\end{tabular}%
}
\caption{Comparison of FOMOS and curvature-based sizing at different resolutions and $\varepsilon$ values.}
\label{tab:sizing_comparison}
\vspace{-0.5cm}
\end{table}

\color{black}
\paragraph{Alternative refinement strategy}
Under a common triangle budget, we compare FOMOS with curvature-based refinement across 50 randomly selected meshes from the McGill dataset. As shown in Table \ref{tab:sizing_comparison}, FOMOS achieves lower Chamfer Distances (CDs) and Hausdorff Distances (HDs) than curvature-based refinement for most tested configurations. At resolution $128^3$ and higher, FOMOS consistently achieves lower CDs and HDs across all tested $\varepsilon$ values. Overall, these results indicate that FOMOS generally improves reconstruction accuracy over curvature-based refinement, particularly at higher resolutions, while reducing worst-case geometric deviations.

\color{black}

\begin{wrapfigure}{R}{\dimexpr1.7in - .95\columnsep\relax}
\vspace{-\intextsep}
\centering
\hspace*{-.95\columnsep}\includegraphics[width=1.5in]{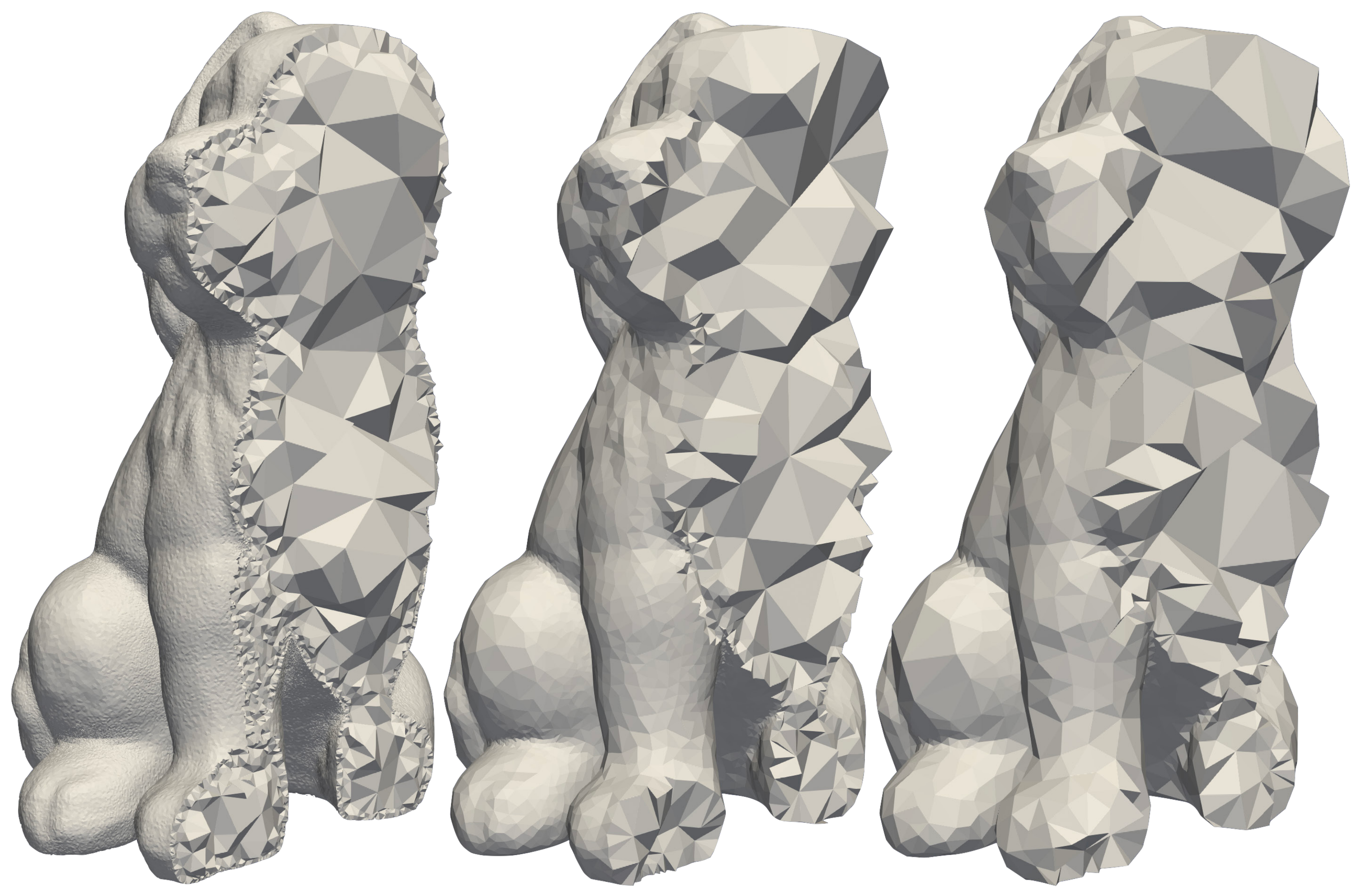}
\vspace{-10pt}
\end{wrapfigure}

\paragraph{Volumetric Meshing. } As the mesher processes a Delaunay tetrahedralization, we can also obtain the volumetric mesh interior to the surface that we control by our $\varepsilon$-condition. For simulations on the fine details of the surface, we do not require details in the interior; however, for the finite element method, the size variation of neighboring elements affects the quality. Therefore, we limit the ratio between the circumradius of a tetrahedron and its shortest edge to 2, which is also the upper theoretical bound that guarantees the Delaunay refinement to terminate \cite{cheng2010delaunay}. See the inset for cuts showing volumetric meshing of various $\varepsilon$ values ($0.1, 0.5, 1.0$).

\paragraph{Application 1: Accelerating FEM Simulations}

\begin{figure}
    \centering
    \begin{subfigure}[t]{0.11\textwidth}
        \includegraphics[width=\textwidth]{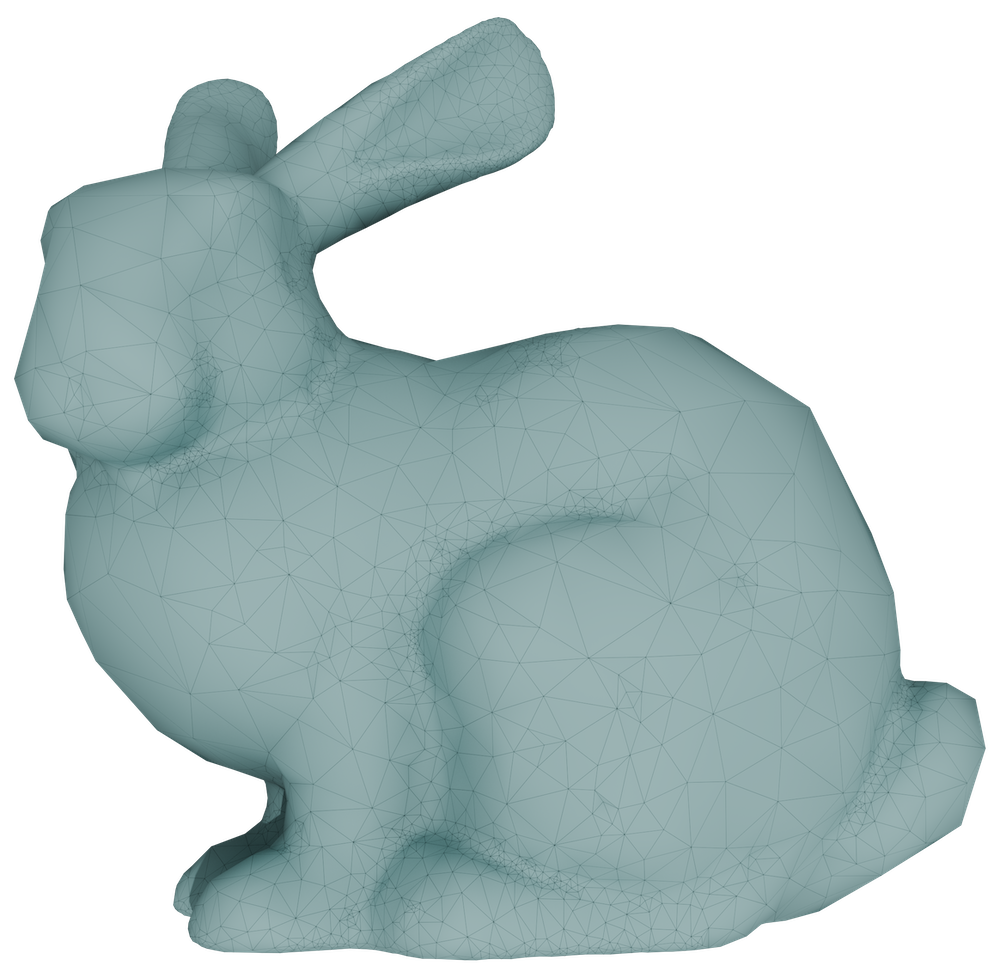}
        \caption{$\varepsilon$=0.5 meshed, 0.00204 min edge}
    \end{subfigure}
    \hfill
    \begin{subfigure}[t]{0.11\textwidth}
        \includegraphics[width=\textwidth]{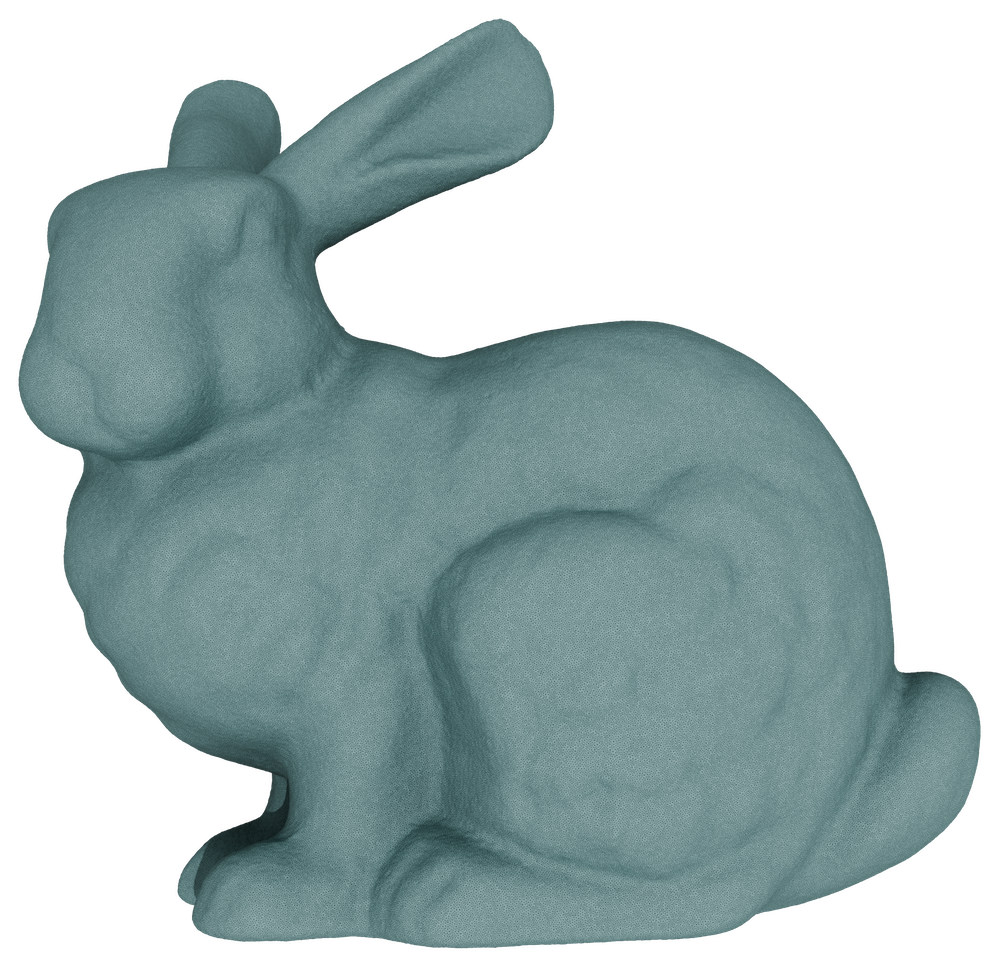}
        \caption{uniform edge length=0.00204}
    \end{subfigure}
    \begin{subfigure}[t]{0.11\textwidth}
        \includegraphics[width=\textwidth]{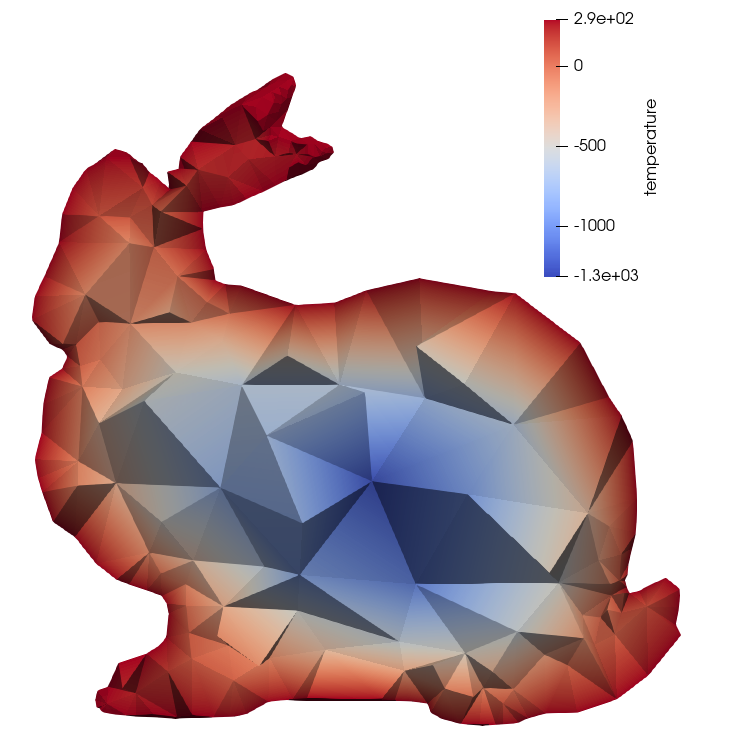}
        \caption{15k tetra, simulation: 0.65s}
    \end{subfigure}
    \hfill
    \begin{subfigure}[t]{0.11\textwidth}
        \includegraphics[width=\textwidth]{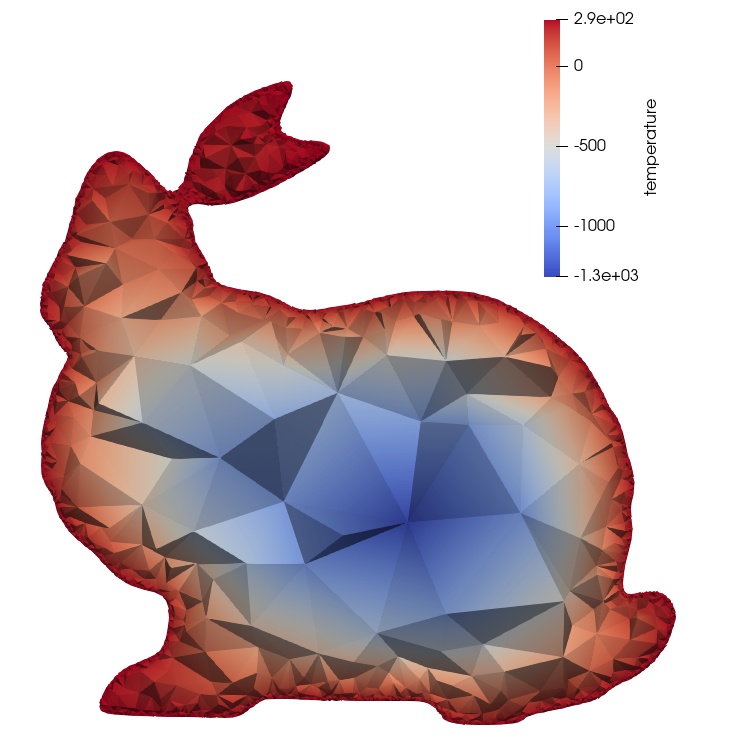}
        \caption{3810k tetra, simulation: 130s}
    \end{subfigure}
  \caption{Left: {\scshape Bunny} first meshed with $\varepsilon=0.5$, then with uniform edge length corresponding to its minimum edge length to ensure equal surface detail. Right: Cuts of corresponding volumetric meshes with heat simulation.}
  \label{fig:heat-simulation}
\end{figure}

\begin{figure*}
    \centering
    \includegraphics[width=\linewidth]{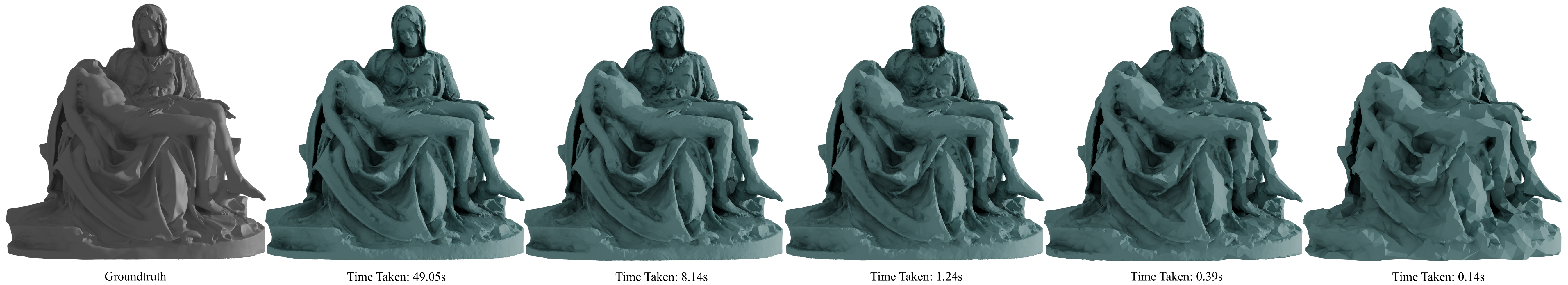}
    \caption{Result of FOMOS on a huge $1024^3$ {\scshape Pieta} input and quick previews for k=1, 2, 4, 8, and 16 ($\varepsilon=1$). The runtime on a standard desktop PC is shown at the bottom. Notably, while Marching Cubes is the only baseline capable of processing an input of this magnitude, it cannot be used for instant previewing.}
    \label{fig:pieta}
\end{figure*}

\begin{table*}[htb]
\centering
\resizebox{\linewidth}{!}{%
\begin{tabular}{llrrrrrrrrrrrrrrrr}
\toprule
\textbf{Resolution} & \textbf{$k$} &
\multicolumn{4}{c}{$\boldsymbol{\varepsilon=0.1}$} &
\multicolumn{4}{c}{$\boldsymbol{\varepsilon=0.2}$} &
\multicolumn{4}{c}{$\boldsymbol{\varepsilon=0.4}$} &
\multicolumn{4}{c}{$\boldsymbol{\varepsilon=0.8}$} \\
\cmidrule(lr){3-6} \cmidrule(lr){7-10}
\cmidrule(lr){11-14} \cmidrule(lr){15-18}
& &
\textbf{Time (s)} & \textbf{Mem (MB)} & \textbf{CD} $\downarrow$ & \textbf{HD} $\downarrow$ &
\textbf{Time (s)} & \textbf{Mem (MB)} & \textbf{CD} $\downarrow$ & \textbf{HD} $\downarrow$ &
\textbf{Time (s)} & \textbf{Mem (MB)} & \textbf{CD} $\downarrow$ & \textbf{HD} $\downarrow$ &
\textbf{Time (s)} & \textbf{Mem (MB)} & \textbf{CD} $\downarrow$ & \textbf{HD} $\downarrow$ \\
\midrule

\multirow{3}{*}{64}
& 1 & 5.008 & 294.609 & $3.84\times10^{-5}$ & 0.0149
& 1.41 & 101.437 & $1.13\times10^{-5}$ & 0.0225
& 0.514 & 45.138 & $4.47\times10^{-4}$ & 0.0387
& 0.351 & 27.376 & $1.17\times10^{-3}$ & 0.0753 \\

& \cellcolor{lightgray}2
& \cellcolor{lightgray}4.394 & \cellcolor{lightgray}254.757
& \cellcolor{lightgray}$3.83\times10^{-5}$ & \cellcolor{lightgray}0.0155
& \cellcolor{lightgray}1.69 & \cellcolor{lightgray}87.266
& \cellcolor{lightgray}$1.93\times10^{-4}$ & \cellcolor{lightgray}0.0274
& \cellcolor{lightgray}0.462 & \cellcolor{lightgray}42.437
& \cellcolor{lightgray}$6.95\times10^{-4}$ & \cellcolor{lightgray}0.0501
& \cellcolor{lightgray}0.303 & \cellcolor{lightgray}23.809
& \cellcolor{lightgray}$1.63\times10^{-3}$ & \cellcolor{lightgray}0.0884 \\

& 4 & 2.992 & 181.376 & $5.75\times10^{-5}$ & 0.019
& 0.818 & 66.393 & $2.80\times10^{-4}$ & 0.0374
& 0.385 & 36.743 & $3.01\times10^{-3}$ & 0.0935
& 0.264 & 21.09 & $6.45\times10^{-3}$ & 0.1925 \\
\midrule

\multirow{4}{*}{128}
& 1 & 7.938 & 460.08 & $1.62\times10^{-5}$ & 0.0092
& 2.168 & 187.886 & $2.53\times10^{-5}$ & 0.0114
& 0.878 & 107.989 & $6.17\times10^{-5}$ & 0.0219
& 0.626 & 87.155 & $2.07\times10^{-4}$ & 0.0425 \\

& \cellcolor{lightgray}2
& \cellcolor{lightgray}6.973 & \cellcolor{lightgray}404.178
& \cellcolor{lightgray}$1.66\times10^{-5}$ & \cellcolor{lightgray}0.0093
& \cellcolor{lightgray}1.927 & \cellcolor{lightgray}166.828
& \cellcolor{lightgray}$2.69\times10^{-5}$ & \cellcolor{lightgray}0.0119
& \cellcolor{lightgray}0.708 & \cellcolor{lightgray}99.98
& \cellcolor{lightgray}$6.86\times10^{-5}$ & \cellcolor{lightgray}0.0233
& \cellcolor{lightgray}0.518 & \cellcolor{lightgray}77.872
& \cellcolor{lightgray}$2.35\times10^{-4}$ & \cellcolor{lightgray}0.0445 \\

& 4 & 6.021 & 328.265 & $1.78\times10^{-5}$ & 0.0094
& 1.432 & 143.584 & $3.19\times10^{-5}$ & 0.0137
& 0.632 & 93.072 & $1.21\times10^{-4}$ & 0.0319
& 0.467 & 71.085 & $6.54\times10^{-4}$ & 0.0713 \\

& \cellcolor{lightgray}8
& \cellcolor{lightgray}3.246 & \cellcolor{lightgray}210.998
& \cellcolor{lightgray}$4.28\times10^{-5}$ & \cellcolor{lightgray}0.0173
& \cellcolor{lightgray}0.915 & \cellcolor{lightgray}111.089
& \cellcolor{lightgray}$2.44\times10^{-4}$ & \cellcolor{lightgray}0.0395
& \cellcolor{lightgray}0.525 & \cellcolor{lightgray}83.975
& \cellcolor{lightgray}$1.43\times10^{-3}$ & \cellcolor{lightgray}0.1235
& \cellcolor{lightgray}0.401 & \cellcolor{lightgray}64.336
& \cellcolor{lightgray}$1.04\times10^{-2}$ & \cellcolor{lightgray}0.2708 \\
\midrule

\multirow{5}{*}{256}
& 1 & 11.43 & 1036.744 & $1.45\times10^{-5}$ & 0.0093
& 4.121 & 650.188 & $2.00\times10^{-5}$ & 0.0099
& 2.007 & 532.004 & $4.17\times10^{-5}$ & 0.0173
& 1.591 & 501.699 & $1.28\times10^{-4}$ & 0.0337 \\

& \cellcolor{lightgray}2
& \cellcolor{lightgray}9.866 & \cellcolor{lightgray}955.963
& \cellcolor{lightgray}$1.48\times10^{-5}$ & \cellcolor{lightgray}0.0092
& \cellcolor{lightgray}3.111 & \cellcolor{lightgray}613.293
& \cellcolor{lightgray}$2.10\times10^{-5}$ & \cellcolor{lightgray}0.0102
& \cellcolor{lightgray}1.375 & \cellcolor{lightgray}506.627
& \cellcolor{lightgray}$4.60\times10^{-5}$ & \cellcolor{lightgray}0.018
& \cellcolor{lightgray}1.007 & \cellcolor{lightgray}478.036
& \cellcolor{lightgray}$1.45\times10^{-4}$ & \cellcolor{lightgray}0.0356 \\

& 4 & 8.034 & 850.44 & $1.54\times10^{-5}$ & 0.0092
& 2.478 & 574.527 & $2.36\times10^{-5}$ & 0.0109
& 1.167 & 493.962 & $5.63\times10^{-5}$ & 0.0208
& 0.909 & 470.49 & $1.86\times10^{-4}$ & 0.0407 \\

& \cellcolor{lightgray}8
& \cellcolor{lightgray}5.754 & \cellcolor{lightgray}720.936
& \cellcolor{lightgray}$1.74\times10^{-5}$ & \cellcolor{lightgray}0.0095
& \cellcolor{lightgray}1.844 & \cellcolor{lightgray}533.689
& \cellcolor{lightgray}$3.17\times10^{-5}$ & \cellcolor{lightgray}0.0149
& \cellcolor{lightgray}1.043 & \cellcolor{lightgray}483.458
& \cellcolor{lightgray}$1.83\times10^{-4}$ & \cellcolor{lightgray}0.0367
& \cellcolor{lightgray}0.847 & \cellcolor{lightgray}460.917
& \cellcolor{lightgray}$4.75\times10^{-3}$ & \cellcolor{lightgray}0.1146 \\

& 16 & 3.061 & 568.176 & $9.12\times10^{-5}$ & 0.0287
& 1.229 & 493.603 & $5.15\times10^{-4}$ & 0.0784
& 0.863 & 467.037 & $1.02\times10^{-2}$ & 0.2298
& 0.754 & 452.313 & $2.77\times10^{-2}$ & 0.3782 \\
\midrule

\multirow{6}{*}{512}
& 1 & 24.935 & 4794.341 & $1.32\times10^{-5}$ & 0.0094
& 12.106 & 4000.459 & $1.57\times10^{-5}$ & 0.0094
& 6.976 & 3783.576 & $2.54\times10^{-5}$ & 0.014
& 5.851 & 3722.541 & $6.37\times10^{-5}$ & 0.0258 \\

& \cellcolor{lightgray}2
& \cellcolor{lightgray}19.817 & \cellcolor{lightgray}4517.346
& \cellcolor{lightgray}$1.34\times10^{-5}$ & \cellcolor{lightgray}0.0093
& \cellcolor{lightgray}7.998 & \cellcolor{lightgray}3878.329
& \cellcolor{lightgray}$1.64\times10^{-5}$ & \cellcolor{lightgray}0.0094
& \cellcolor{lightgray}4.579 & \cellcolor{lightgray}3686.458
& \cellcolor{lightgray}$2.83\times10^{-5}$ & \cellcolor{lightgray}0.0146
& \cellcolor{lightgray}3.843 & \cellcolor{lightgray}3631.752
& \cellcolor{lightgray}$7.51\times10^{-5}$ & \cellcolor{lightgray}0.0275 \\

& 4 & 14.408 & 4240.037 & $1.39\times10^{-5}$ & 0.0092
& 6.917 & 3792.967 & $1.82\times10^{-5}$ & 0.0096
& 3.944 & 3652.863 & $3.56\times10^{-5}$ & 0.0163
& 3.49 & 3616.672 & $1.04\times10^{-4}$ & 0.0306 \\

& \cellcolor{lightgray}8
& \cellcolor{lightgray}10.459 & \cellcolor{lightgray}3993.843
& \cellcolor{lightgray}$1.51\times10^{-5}$ & \cellcolor{lightgray}0.0091
& \cellcolor{lightgray}5.053 & \cellcolor{lightgray}3717.388
& \cellcolor{lightgray}$2.27\times10^{-5}$ & \cellcolor{lightgray}0.0113
& \cellcolor{lightgray}3.698 & \cellcolor{lightgray}3631.625
& \cellcolor{lightgray}$5.33\times10^{-5}$ & \cellcolor{lightgray}0.0211
& \cellcolor{lightgray}3.442 & \cellcolor{lightgray}3608.285
& \cellcolor{lightgray}$2.27\times10^{-4}$ & \cellcolor{lightgray}0.0439 \\

& 16 & 7.356 & 3806.176 & $1.96\times10^{-5}$ & 0.0114
& 4.104 & 3656.503 & $4.53\times10^{-5}$ & 0.0253
& 3.474 & 3617.255 & $6.85\times10^{-4}$ & 0.0851
& 3.334 & 3594.713 & $9.16\times10^{-3}$ & 0.2154 \\

& \cellcolor{lightgray}32
& \cellcolor{lightgray}4.873 & \cellcolor{lightgray}3663.176
& \cellcolor{lightgray}$2.40\times10^{-4}$ & \cellcolor{lightgray}0.0553
& \cellcolor{lightgray}3.555 & \cellcolor{lightgray}3618.658
& \cellcolor{lightgray}$2.00\times10^{-3}$ & \cellcolor{lightgray}0.1588
& \cellcolor{lightgray}3.311 & \cellcolor{lightgray}3596.977
& \cellcolor{lightgray}$1.26\times10^{-2}$ & \cellcolor{lightgray}0.3059
& \cellcolor{lightgray}3.303 & \cellcolor{lightgray}3587.161
& \cellcolor{lightgray}$3.54\times10^{-2}$ & \cellcolor{lightgray}0.4333 \\
\bottomrule
\end{tabular}%
}
\caption{\textcolor{black}{Runtime, memory consumption, Chamfer Distance (CD), and Hausdorff Distance (HD) for different $k$ values, under varying resolutions and $\varepsilon$ values.}}
\label{tab:k_parameter_evaluation}
\end{table*}

As an application, we demonstrate that a heat simulation on a volumetric mesh (see Figure \ref{fig:heat-simulation}) induced from an $\varepsilon$-sampled surface mesh generated by our method has many fewer elements and can therefore perform 200x faster while capturing the same level of detail. 
As we construct the mesh directly from an SDF, which can also be deformed interactively, e.g., using VDB \cite{museth2013vdb}, we can reconstruct and visualize meshes from CFDs without requiring any intermediate meshes.

\paragraph{Application 2: Instant SDF previewing} While meshing a large SDF, e.g., $512^3$ is relatively fast, it still takes several seconds. FOMOS can substantially reduce it by using a subsampled medial scaffold during meshing of the entire SDF (\textit{Note that we subsample only the MA, not the SDF}).
Figure \ref{fig:pieta} shows an example where subsampling reduces the original runtime from 49s down to 0.14s for a factor of 16 ($k=16$)
, approximately quadratic with the given subsampling factor. This allows quick SDF previewing for inspection or thumbnail creation.

\color{black}
 To investigate the effect of subsampling ($k$) on runtime, memory consumption, and reconstruction accuracy, we evaluated FOMOS on randomly selected 50 meshes from the McGill dataset. As shown in Table~\ref{tab:k_parameter_evaluation}, increasing $k$ generally reduces both runtime and memory consumption, at the cost of increased reconstruction error. For example, at a resolution of $512^3$ with $\varepsilon=0.1$, increasing $k$ from 1 to 32 reduces runtime by 80.5\% and memory consumption by 23.6\%. The impact on reconstruction accuracy is more pronounced at larger values of $\varepsilon$. At the same resolution with $\varepsilon=0.8$, increasing $k$ from 1 to 32 increases the Chamfer Distance from $6.37\times10^{-5}$ to $3.54\times10^{-2}$ and the Hausdorff Distance from 0.0258 to 0.4333. In contrast, at smaller values of $\varepsilon$, moderate increases in $k$ can yield substantial computational savings while maintaining relatively low reconstruction errors. Overall, $k$ provides a means of balancing computational efficiency against geometric fidelity, with smaller values generally favoring accuracy and larger values favoring efficiency.

\color{black}

\color{black}

\section{Limitations \& Future Work}

\begin{figure}
    \centering
    \includegraphics[width=1.0\linewidth]{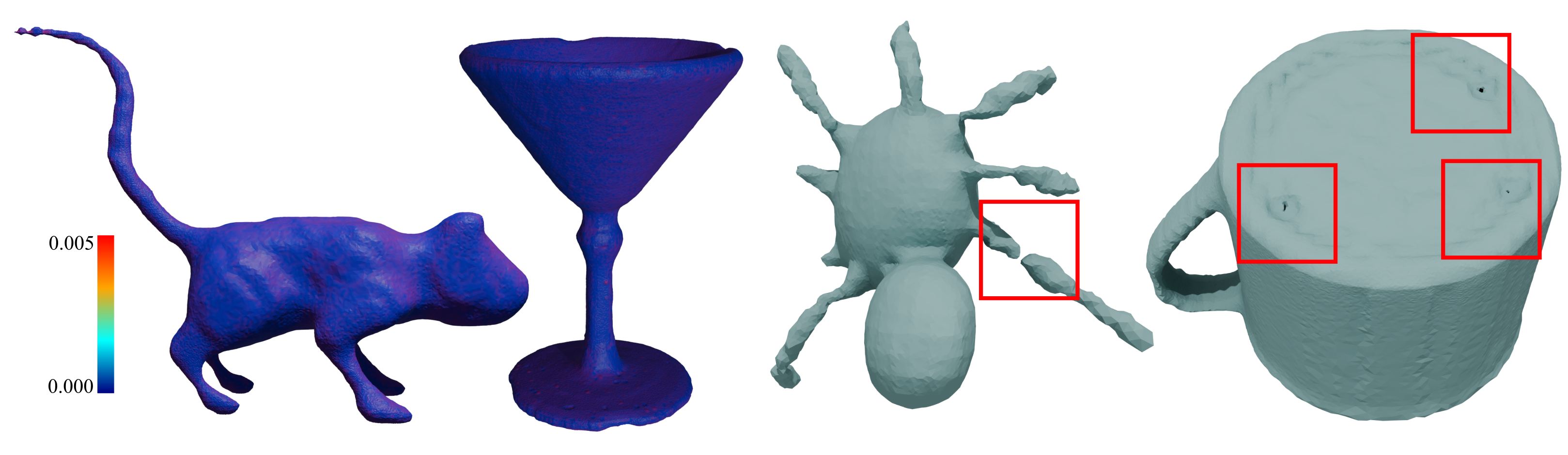}
    \caption{Left: Reconstructions from the dataset with the highest Chamfer Distance (colored based on GT distance) on $128^3$ (CD = $4.7\times 10^{-4}$ and $1.5 \times 10^{-4}$), due to thin features and surface noise. Right: Two $32^3$ models where FOMOS reconstructs incorrect topologies, resulting in mismatched 0th+1st Betti numbers from thin features and undersampling.}
    \label{faulty}
\end{figure}

Despite its efficiency in terms of running time, memory requirement, and accuracy, FOMOS has a few limitations:

\paragraph{Resolution constraints} As illustrated in various examples, FOMOS is not designed for reconstruction from very low-resolution inputs. 
Using medial balls to beautify the surfaces is an interesting direction.
\paragraph{Capturing sharp features} As in any other SOTA methods, the reconstructed mesh may exhibit jagged edges along sharp features such as creases due to the Delaunay property of the underlying mesher. This could easily be fixed using a post-processing algorithm such as EdgeSharpener \cite{attene2003edge} to recover sharp features.
\paragraph{Capturing thin features} Accurately reconstructing thin features requires high grid resolution. As shown in Figure \ref{faulty}, insufficient sampling density causes thin features to deteriorate or disconnect.


\paragraph{Future work} 
We will reduce sphere-fitting iterations by bounding maximum spheres faster than bisection using closest points; however, this requires increasing the smoothness of the zero surface, e.g., by triquadratic or tricubic interpolation of SDF grid values.
Although we showed results from SDFs sampled on grids, our method is not dependent on those and can also handle SDFs that answer queries on arbitrary underlying representations, including neural SDFs, such as DeepSDF \cite{park2019deepsdf}, or spatial hierarchies. Our medial point computation should also parallelize well on the GPU.

\section{Conclusion}

In this work, we presented an end-to-end pipeline for efficiently meshing arbitrary SDFs with triangles proportional to local features, thereby directly and quickly creating concise representations even for huge datasets. These representations possess properties that make them highly suitable for simulations and are fast enough for interactively deforming SDFs, such as those from CFDs, thanks to our simple and efficiently computed medial scaffolds. Additionally, computing the medial scaffold on a sparse subset of the grid allows the reconstruction of the full SDF, enabling instant previews, as runtime becomes output-sensitive rather than grid-dependent.





\begin{acks}
This work has been partially funded by the Wiener Wissenschafts-, Forschungs- und Technologiefonds (WWTF) project ICT19-009, FWF project PAT9402024, the ANR JCJC project SketchMAD (ANR-23-CE33-0009), and generous support from Adobe.

\end{acks}

\bibliographystyle{ACM-Reference-Format}
\bibliography{sample-base}

\end{document}